\documentclass[longauth]{aa}
\usepackage{widetext}

\usepackage{graphicx}
\usepackage{natbib}
\usepackage{scalerel}
\usepackage{dcolumn}
\usepackage{siunitx}
\usepackage{multirow}
\usepackage{booktabs}
\usepackage{bm}

\usepackage[table]{xcolor}
\usepackage{txfonts}
\usepackage[pdfencoding=auto,psdextra]{hyperref}
\hypersetup{
    colorlinks=true,
    linkcolor=blue,
    filecolor=magenta,      
    urlcolor=blue,
    citecolor=blue
}
\makeatletter
\renewcommand*\aa@pageof{, page \thepage{} of \pageref*{LastPage}}
\makeatother

\usepackage[utf8]{inputenc}

\usepackage[switch, modulo]{lineno}

\usepackage{euclid}
\usepackage{orcidlink}

\newcommand{\Msun}{M_\sun}
\newcommand{\Mhalo}{M_\mathrm{h}}
\newcommand{\logMhalo}{\logten(\Mhalo/\Msun)}
\newcommand{\Mstar}{M_*}
\newcommand{\logMstar}{\logten(\Mstar/\Msun)}
\newcommand{\fICL}{f_\textrm{ICL}}
\newcommand{\Rmax}{R_\textrm{max}}
\newcommand{\SN}{\textrm{S/N}}
\newcommand{\ICLap}{(BCG+ICL)$_{{50}-{200}}$}

\newcommand{\eceb}[1]{{#1}}

\newcommand{\referee}[1]{#1}

\begin{document}
%
%
    \title{\Euclid preparation}
    \subtitle{LXX. Forecasting detection limits for intracluster light in the Euclid Wide Survey}

\newcommand{\orcid}[1]{\orcidlink{#1}} 
\author{Euclid Collaboration: C.~Bellhouse\orcid{0000-0002-6179-8007}\thanks{\email{callum.bellhouse@nottingham.ac.uk}}\inst{\ref{aff1}}
\and J.~B.~Golden-Marx\orcid{0000-0002-6394-045X}\inst{\ref{aff1}}
\and S.~P.~Bamford\orcid{0000-0001-7821-7195}\inst{\ref{aff1}}
\and N.~A.~Hatch\orcid{0000-0001-5600-0534}\inst{\ref{aff1}}
\and M.~Kluge\orcid{0000-0002-9618-2552}\inst{\ref{aff2}}
\and A.~Ellien\orcid{0000-0002-1038-3370}\inst{\ref{aff3}}
\and S.~L.~Ahad\orcid{0000-0001-6336-642X}\inst{\ref{aff4},\ref{aff5}}
\and P.~Dimauro\orcid{0000-0001-7399-2854}\inst{\ref{aff6},\ref{aff7}}
\and F.~Durret\orcid{0000-0002-6991-4578}\inst{\ref{aff8}}
\and A.~H.~Gonzalez\orcid{0000-0002-0933-8601}\inst{\ref{aff9}}
\and Y.~Jimenez-Teja\orcid{0000-0002-6090-2853}\inst{\ref{aff10},\ref{aff7}}
\and M.~Montes\orcid{0000-0001-7847-0393}\inst{\ref{aff11},\ref{aff12},\ref{aff13}}
\and M.~Sereno\orcid{0000-0003-0302-0325}\inst{\ref{aff14},\ref{aff15}}
\and E.~Slezak\orcid{0000-0003-4771-7263}\inst{\ref{aff16}}
\and M.~Bolzonella\orcid{0000-0003-3278-4607}\inst{\ref{aff14}}
\and G.~Castignani\orcid{0000-0001-6831-0687}\inst{\ref{aff14}}
\and O.~Cucciati\orcid{0000-0002-9336-7551}\inst{\ref{aff14}}
\and G.~De~Lucia\orcid{0000-0002-6220-9104}\inst{\ref{aff17}}
\and Z.~Ghaffari\orcid{0000-0002-6467-8078}\inst{\ref{aff17},\ref{aff18}}
\and L.~Moscardini\orcid{0000-0002-3473-6716}\inst{\ref{aff19},\ref{aff14},\ref{aff15}}
\and R.~Pello\orcid{0000-0003-0858-6109}\inst{\ref{aff20}}
\and L.~Pozzetti\orcid{0000-0001-7085-0412}\inst{\ref{aff14}}
\and T.~Saifollahi\orcid{0000-0002-9554-7660}\inst{\ref{aff21}}
\and A.~S.~Borlaff\orcid{0000-0003-3249-4431}\inst{\ref{aff22},\ref{aff23}}
\and N.~Aghanim\orcid{0000-0002-6688-8992}\inst{\ref{aff24}}
\and B.~Altieri\orcid{0000-0003-3936-0284}\inst{\ref{aff25}}
\and A.~Amara\inst{\ref{aff26}}
\and S.~Andreon\orcid{0000-0002-2041-8784}\inst{\ref{aff27}}
\and C.~Baccigalupi\orcid{0000-0002-8211-1630}\inst{\ref{aff18},\ref{aff17},\ref{aff28},\ref{aff29}}
\and M.~Baldi\orcid{0000-0003-4145-1943}\inst{\ref{aff30},\ref{aff14},\ref{aff15}}
\and S.~Bardelli\orcid{0000-0002-8900-0298}\inst{\ref{aff14}}
\and A.~Basset\inst{\ref{aff31}}
\and P.~Battaglia\orcid{0000-0002-7337-5909}\inst{\ref{aff14}}
\and R.~Bender\orcid{0000-0001-7179-0626}\inst{\ref{aff2},\ref{aff32}}
\and D.~Bonino\orcid{0000-0002-3336-9977}\inst{\ref{aff33}}
\and E.~Branchini\orcid{0000-0002-0808-6908}\inst{\ref{aff34},\ref{aff35},\ref{aff27}}
\and M.~Brescia\orcid{0000-0001-9506-5680}\inst{\ref{aff36},\ref{aff37},\ref{aff38}}
\and A.~Caillat\inst{\ref{aff20}}
\and S.~Camera\orcid{0000-0003-3399-3574}\inst{\ref{aff39},\ref{aff40},\ref{aff33}}
\and V.~Capobianco\orcid{0000-0002-3309-7692}\inst{\ref{aff33}}
\and C.~Carbone\orcid{0000-0003-0125-3563}\inst{\ref{aff41}}
\and V.~F.~Cardone\inst{\ref{aff6},\ref{aff42}}
\and J.~Carretero\orcid{0000-0002-3130-0204}\inst{\ref{aff43},\ref{aff44}}
\and S.~Casas\orcid{0000-0002-4751-5138}\inst{\ref{aff45},\ref{aff46}}
\and M.~Castellano\orcid{0000-0001-9875-8263}\inst{\ref{aff6}}
\and S.~Cavuoti\orcid{0000-0002-3787-4196}\inst{\ref{aff37},\ref{aff38}}
\and A.~Cimatti\inst{\ref{aff47}}
\and C.~Colodro-Conde\inst{\ref{aff12}}
\and G.~Congedo\orcid{0000-0003-2508-0046}\inst{\ref{aff48}}
\and C.~J.~Conselice\orcid{0000-0003-1949-7638}\inst{\ref{aff49}}
\and L.~Conversi\orcid{0000-0002-6710-8476}\inst{\ref{aff50},\ref{aff25}}
\and Y.~Copin\orcid{0000-0002-5317-7518}\inst{\ref{aff51}}
\and F.~Courbin\orcid{0000-0003-0758-6510}\inst{\ref{aff52},\ref{aff53}}
\and H.~M.~Courtois\orcid{0000-0003-0509-1776}\inst{\ref{aff54}}
\and J.-C.~Cuillandre\orcid{0000-0002-3263-8645}\inst{\ref{aff55}}
\and A.~Da~Silva\orcid{0000-0002-6385-1609}\inst{\ref{aff56},\ref{aff57}}
\and H.~Degaudenzi\orcid{0000-0002-5887-6799}\inst{\ref{aff58}}
\and A.~M.~Di~Giorgio\orcid{0000-0002-4767-2360}\inst{\ref{aff59}}
\and J.~Dinis\orcid{0000-0001-5075-1601}\inst{\ref{aff56},\ref{aff57}}
\and F.~Dubath\orcid{0000-0002-6533-2810}\inst{\ref{aff58}}
\and C.~A.~J.~Duncan\orcid{0009-0003-3573-0791}\inst{\ref{aff49}}
\and X.~Dupac\inst{\ref{aff25}}
\and S.~Dusini\orcid{0000-0002-1128-0664}\inst{\ref{aff60}}
\and M.~Farina\orcid{0000-0002-3089-7846}\inst{\ref{aff59}}
\and S.~Farrens\orcid{0000-0002-9594-9387}\inst{\ref{aff55}}
\and F.~Faustini\orcid{0000-0001-6274-5145}\inst{\ref{aff61},\ref{aff6}}
\and S.~Ferriol\inst{\ref{aff51}}
\and S.~Fotopoulou\orcid{0000-0002-9686-254X}\inst{\ref{aff62}}
\and M.~Frailis\orcid{0000-0002-7400-2135}\inst{\ref{aff17}}
\and E.~Franceschi\orcid{0000-0002-0585-6591}\inst{\ref{aff14}}
\and M.~Fumana\orcid{0000-0001-6787-5950}\inst{\ref{aff41}}
\and S.~Galeotta\orcid{0000-0002-3748-5115}\inst{\ref{aff17}}
\and K.~George\orcid{0000-0002-1734-8455}\inst{\ref{aff32}}
\and B.~Gillis\orcid{0000-0002-4478-1270}\inst{\ref{aff48}}
\and C.~Giocoli\orcid{0000-0002-9590-7961}\inst{\ref{aff14},\ref{aff15}}
\and P.~G\'omez-Alvarez\orcid{0000-0002-8594-5358}\inst{\ref{aff63},\ref{aff25}}
\and A.~Grazian\orcid{0000-0002-5688-0663}\inst{\ref{aff64}}
\and F.~Grupp\inst{\ref{aff2},\ref{aff32}}
\and S.~V.~H.~Haugan\orcid{0000-0001-9648-7260}\inst{\ref{aff65}}
\and H.~Hoekstra\orcid{0000-0002-0641-3231}\inst{\ref{aff66}}
\and M.~S.~Holliman\inst{\ref{aff48}}
\and W.~Holmes\inst{\ref{aff67}}
\and I.~Hook\orcid{0000-0002-2960-978X}\inst{\ref{aff68}}
\and F.~Hormuth\inst{\ref{aff69}}
\and A.~Hornstrup\orcid{0000-0002-3363-0936}\inst{\ref{aff70},\ref{aff71}}
\and P.~Hudelot\inst{\ref{aff72}}
\and K.~Jahnke\orcid{0000-0003-3804-2137}\inst{\ref{aff73}}
\and M.~Jhabvala\inst{\ref{aff74}}
\and E.~Keih\"anen\orcid{0000-0003-1804-7715}\inst{\ref{aff75}}
\and S.~Kermiche\orcid{0000-0002-0302-5735}\inst{\ref{aff76}}
\and A.~Kiessling\orcid{0000-0002-2590-1273}\inst{\ref{aff67}}
\and M.~Kilbinger\orcid{0000-0001-9513-7138}\inst{\ref{aff55}}
\and B.~Kubik\orcid{0009-0006-5823-4880}\inst{\ref{aff51}}
\and M.~K\"ummel\orcid{0000-0003-2791-2117}\inst{\ref{aff32}}
\and M.~Kunz\orcid{0000-0002-3052-7394}\inst{\ref{aff77}}
\and H.~Kurki-Suonio\orcid{0000-0002-4618-3063}\inst{\ref{aff78},\ref{aff79}}
\and P.~Liebing\inst{\ref{aff80}}
\and S.~Ligori\orcid{0000-0003-4172-4606}\inst{\ref{aff33}}
\and P.~B.~Lilje\orcid{0000-0003-4324-7794}\inst{\ref{aff65}}
\and V.~Lindholm\orcid{0000-0003-2317-5471}\inst{\ref{aff78},\ref{aff79}}
\and I.~Lloro\orcid{0000-0001-5966-1434}\inst{\ref{aff81}}
\and G.~Mainetti\orcid{0000-0003-2384-2377}\inst{\ref{aff82}}
\and D.~Maino\inst{\ref{aff83},\ref{aff41},\ref{aff84}}
\and E.~Maiorano\orcid{0000-0003-2593-4355}\inst{\ref{aff14}}
\and O.~Mansutti\orcid{0000-0001-5758-4658}\inst{\ref{aff17}}
\and O.~Marggraf\orcid{0000-0001-7242-3852}\inst{\ref{aff85}}
\and K.~Markovic\orcid{0000-0001-6764-073X}\inst{\ref{aff67}}
\and M.~Martinelli\orcid{0000-0002-6943-7732}\inst{\ref{aff6},\ref{aff42}}
\and N.~Martinet\orcid{0000-0003-2786-7790}\inst{\ref{aff20}}
\and F.~Marulli\orcid{0000-0002-8850-0303}\inst{\ref{aff19},\ref{aff14},\ref{aff15}}
\and R.~Massey\orcid{0000-0002-6085-3780}\inst{\ref{aff86}}
\and S.~Maurogordato\inst{\ref{aff16}}
\and E.~Medinaceli\orcid{0000-0002-4040-7783}\inst{\ref{aff14}}
\and S.~Mei\orcid{0000-0002-2849-559X}\inst{\ref{aff87}}
\and M.~Melchior\inst{\ref{aff88}}
\and M.~Meneghetti\orcid{0000-0003-1225-7084}\inst{\ref{aff14},\ref{aff15}}
\and E.~Merlin\orcid{0000-0001-6870-8900}\inst{\ref{aff6}}
\and G.~Meylan\inst{\ref{aff89}}
\and M.~Moresco\orcid{0000-0002-7616-7136}\inst{\ref{aff19},\ref{aff14}}
\and R.~Nakajima\orcid{0009-0009-1213-7040}\inst{\ref{aff85}}
\and C.~Neissner\orcid{0000-0001-8524-4968}\inst{\ref{aff90},\ref{aff44}}
\and S.-M.~Niemi\inst{\ref{aff91}}
\and C.~Padilla\orcid{0000-0001-7951-0166}\inst{\ref{aff90}}
\and S.~Paltani\orcid{0000-0002-8108-9179}\inst{\ref{aff58}}
\and F.~Pasian\orcid{0000-0002-4869-3227}\inst{\ref{aff17}}
\and K.~Pedersen\inst{\ref{aff92}}
\and V.~Pettorino\inst{\ref{aff91}}
\and S.~Pires\orcid{0000-0002-0249-2104}\inst{\ref{aff55}}
\and G.~Polenta\orcid{0000-0003-4067-9196}\inst{\ref{aff61}}
\and M.~Poncet\inst{\ref{aff31}}
\and L.~A.~Popa\inst{\ref{aff93}}
\and F.~Raison\orcid{0000-0002-7819-6918}\inst{\ref{aff2}}
\and A.~Renzi\orcid{0000-0001-9856-1970}\inst{\ref{aff94},\ref{aff60}}
\and J.~Rhodes\orcid{0000-0002-4485-8549}\inst{\ref{aff67}}
\and G.~Riccio\inst{\ref{aff37}}
\and E.~Romelli\orcid{0000-0003-3069-9222}\inst{\ref{aff17}}
\and M.~Roncarelli\orcid{0000-0001-9587-7822}\inst{\ref{aff14}}
\and E.~Rossetti\orcid{0000-0003-0238-4047}\inst{\ref{aff30}}
\and R.~Saglia\orcid{0000-0003-0378-7032}\inst{\ref{aff32},\ref{aff2}}
\and Z.~Sakr\orcid{0000-0002-4823-3757}\inst{\ref{aff95},\ref{aff96},\ref{aff97}}
\and D.~Sapone\orcid{0000-0001-7089-4503}\inst{\ref{aff98}}
\and B.~Sartoris\orcid{0000-0003-1337-5269}\inst{\ref{aff32},\ref{aff17}}
\and P.~Schneider\orcid{0000-0001-8561-2679}\inst{\ref{aff85}}
\and T.~Schrabback\orcid{0000-0002-6987-7834}\inst{\ref{aff99}}
\and G.~Seidel\orcid{0000-0003-2907-353X}\inst{\ref{aff73}}
\and S.~Serrano\orcid{0000-0002-0211-2861}\inst{\ref{aff100},\ref{aff101},\ref{aff11}}
\and C.~Sirignano\orcid{0000-0002-0995-7146}\inst{\ref{aff94},\ref{aff60}}
\and G.~Sirri\orcid{0000-0003-2626-2853}\inst{\ref{aff15}}
\and L.~Stanco\orcid{0000-0002-9706-5104}\inst{\ref{aff60}}
\and J.~Steinwagner\orcid{0000-0001-7443-1047}\inst{\ref{aff2}}
\and P.~Tallada-Cresp\'{i}\orcid{0000-0002-1336-8328}\inst{\ref{aff43},\ref{aff44}}
\and I.~Tereno\inst{\ref{aff56},\ref{aff102}}
\and R.~Toledo-Moreo\orcid{0000-0002-2997-4859}\inst{\ref{aff103}}
\and F.~Torradeflot\orcid{0000-0003-1160-1517}\inst{\ref{aff44},\ref{aff43}}
\and A.~Tsyganov\inst{\ref{aff104}}
\and I.~Tutusaus\orcid{0000-0002-3199-0399}\inst{\ref{aff96}}
\and L.~Valenziano\orcid{0000-0002-1170-0104}\inst{\ref{aff14},\ref{aff105}}
\and T.~Vassallo\orcid{0000-0001-6512-6358}\inst{\ref{aff32},\ref{aff17}}
\and G.~Verdoes~Kleijn\orcid{0000-0001-5803-2580}\inst{\ref{aff106}}
\and A.~Veropalumbo\orcid{0000-0003-2387-1194}\inst{\ref{aff27},\ref{aff35},\ref{aff34}}
\and Y.~Wang\orcid{0000-0002-4749-2984}\inst{\ref{aff107}}
\and J.~Weller\orcid{0000-0002-8282-2010}\inst{\ref{aff32},\ref{aff2}}
\and G.~Zamorani\orcid{0000-0002-2318-301X}\inst{\ref{aff14}}
\and E.~Zucca\orcid{0000-0002-5845-8132}\inst{\ref{aff14}}
\and A.~Biviano\orcid{0000-0002-0857-0732}\inst{\ref{aff17},\ref{aff18}}
\and E.~Bozzo\orcid{0000-0002-8201-1525}\inst{\ref{aff58}}
\and C.~Burigana\orcid{0000-0002-3005-5796}\inst{\ref{aff108},\ref{aff105}}
\and M.~Calabrese\orcid{0000-0002-2637-2422}\inst{\ref{aff109},\ref{aff41}}
\and D.~Di~Ferdinando\inst{\ref{aff15}}
\and J.~A.~Escartin~Vigo\inst{\ref{aff2}}
\and R.~Farinelli\inst{\ref{aff14}}
\and F.~Finelli\orcid{0000-0002-6694-3269}\inst{\ref{aff14},\ref{aff105}}
\and L.~Gabarra\orcid{0000-0002-8486-8856}\inst{\ref{aff110}}
\and J.~Gracia-Carpio\inst{\ref{aff2}}
\and S.~Matthew\orcid{0000-0001-8448-1697}\inst{\ref{aff48}}
\and N.~Mauri\orcid{0000-0001-8196-1548}\inst{\ref{aff47},\ref{aff15}}
\and A.~Mora\orcid{0000-0002-1922-8529}\inst{\ref{aff111}}
\and M.~P\"ontinen\orcid{0000-0001-5442-2530}\inst{\ref{aff78}}
\and V.~Scottez\inst{\ref{aff8},\ref{aff112}}
\and P.~Simon\inst{\ref{aff85}}
\and M.~Tenti\orcid{0000-0002-4254-5901}\inst{\ref{aff15}}
\and M.~Viel\orcid{0000-0002-2642-5707}\inst{\ref{aff18},\ref{aff17},\ref{aff29},\ref{aff28},\ref{aff113}}
\and M.~Wiesmann\orcid{0009-0000-8199-5860}\inst{\ref{aff65}}
\and Y.~Akrami\orcid{0000-0002-2407-7956}\inst{\ref{aff114},\ref{aff115}}
\and I.~T.~Andika\orcid{0000-0001-6102-9526}\inst{\ref{aff116},\ref{aff117}}
\and S.~Anselmi\orcid{0000-0002-3579-9583}\inst{\ref{aff60},\ref{aff94},\ref{aff118}}
\and M.~Archidiacono\orcid{0000-0003-4952-9012}\inst{\ref{aff83},\ref{aff84}}
\and F.~Atrio-Barandela\orcid{0000-0002-2130-2513}\inst{\ref{aff119}}
\and M.~Ballardini\orcid{0000-0003-4481-3559}\inst{\ref{aff120},\ref{aff14},\ref{aff121}}
\and M.~Bethermin\orcid{0000-0002-3915-2015}\inst{\ref{aff21}}
\and A.~Blanchard\orcid{0000-0001-8555-9003}\inst{\ref{aff96}}
\and L.~Blot\orcid{0000-0002-9622-7167}\inst{\ref{aff122},\ref{aff118}}
\and H.~B\"ohringer\orcid{0000-0001-8241-4204}\inst{\ref{aff2},\ref{aff123},\ref{aff124}}
\and S.~Borgani\orcid{0000-0001-6151-6439}\inst{\ref{aff125},\ref{aff18},\ref{aff17},\ref{aff28},\ref{aff113}}
\and M.~L.~Brown\orcid{0000-0002-0370-8077}\inst{\ref{aff49}}
\and S.~Bruton\orcid{0000-0002-6503-5218}\inst{\ref{aff126}}
\and R.~Cabanac\orcid{0000-0001-6679-2600}\inst{\ref{aff96}}
\and A.~Calabro\orcid{0000-0003-2536-1614}\inst{\ref{aff6}}
\and G.~Ca\~nas-Herrera\orcid{0000-0003-2796-2149}\inst{\ref{aff91},\ref{aff127}}
\and A.~Cappi\inst{\ref{aff14},\ref{aff16}}
\and F.~Caro\inst{\ref{aff6}}
\and C.~S.~Carvalho\inst{\ref{aff102}}
\and T.~Castro\orcid{0000-0002-6292-3228}\inst{\ref{aff17},\ref{aff28},\ref{aff18},\ref{aff113}}
\and K.~C.~Chambers\orcid{0000-0001-6965-7789}\inst{\ref{aff128}}
\and F.~Cogato\orcid{0000-0003-4632-6113}\inst{\ref{aff19},\ref{aff14}}
\and T.~Contini\orcid{0000-0003-0275-938X}\inst{\ref{aff96}}
\and A.~R.~Cooray\orcid{0000-0002-3892-0190}\inst{\ref{aff129}}
\and F.~De~Paolis\orcid{0000-0001-6460-7563}\inst{\ref{aff130},\ref{aff131},\ref{aff132}}
\and G.~Desprez\orcid{0000-0001-8325-1742}\inst{\ref{aff106}}
\and A.~D\'iaz-S\'anchez\orcid{0000-0003-0748-4768}\inst{\ref{aff133}}
\and J.~J.~Diaz\inst{\ref{aff134}}
\and S.~Di~Domizio\orcid{0000-0003-2863-5895}\inst{\ref{aff34},\ref{aff35}}
\and J.~M.~Diego\orcid{0000-0001-9065-3926}\inst{\ref{aff135}}
\and H.~Dole\orcid{0000-0002-9767-3839}\inst{\ref{aff24}}
\and S.~Escoffier\orcid{0000-0002-2847-7498}\inst{\ref{aff76}}
\and A.~G.~Ferrari\orcid{0009-0005-5266-4110}\inst{\ref{aff15}}
\and P.~G.~Ferreira\orcid{0000-0002-3021-2851}\inst{\ref{aff110}}
\and A.~Finoguenov\orcid{0000-0002-4606-5403}\inst{\ref{aff78}}
\and A.~Fontana\orcid{0000-0003-3820-2823}\inst{\ref{aff6}}
\and K.~Ganga\orcid{0000-0001-8159-8208}\inst{\ref{aff87}}
\and J.~Garc\'ia-Bellido\orcid{0000-0002-9370-8360}\inst{\ref{aff114}}
\and T.~Gasparetto\orcid{0000-0002-7913-4866}\inst{\ref{aff17}}
\and E.~Gaztanaga\orcid{0000-0001-9632-0815}\inst{\ref{aff11},\ref{aff100},\ref{aff46}}
\and F.~Giacomini\orcid{0000-0002-3129-2814}\inst{\ref{aff15}}
\and F.~Gianotti\orcid{0000-0003-4666-119X}\inst{\ref{aff14}}
\and G.~Gozaliasl\orcid{0000-0002-0236-919X}\inst{\ref{aff136},\ref{aff78}}
\and A.~Gregorio\orcid{0000-0003-4028-8785}\inst{\ref{aff125},\ref{aff17},\ref{aff28}}
\and M.~Guidi\orcid{0000-0001-9408-1101}\inst{\ref{aff30},\ref{aff14}}
\and C.~M.~Gutierrez\orcid{0000-0001-7854-783X}\inst{\ref{aff137}}
\and A.~Hall\orcid{0000-0002-3139-8651}\inst{\ref{aff48}}
\and W.~G.~Hartley\inst{\ref{aff58}}
\and S.~Hemmati\orcid{0000-0003-2226-5395}\inst{\ref{aff138}}
\and H.~Hildebrandt\orcid{0000-0002-9814-3338}\inst{\ref{aff139}}
\and J.~Hjorth\orcid{0000-0002-4571-2306}\inst{\ref{aff92}}
\and A.~Jimenez~Mu\~noz\orcid{0009-0004-5252-185X}\inst{\ref{aff140}}
\and J.~J.~E.~Kajava\orcid{0000-0002-3010-8333}\inst{\ref{aff141},\ref{aff142}}
\and Y.~Kang\orcid{0009-0000-8588-7250}\inst{\ref{aff58}}
\and V.~Kansal\orcid{0000-0002-4008-6078}\inst{\ref{aff143},\ref{aff144}}
\and D.~Karagiannis\orcid{0000-0002-4927-0816}\inst{\ref{aff120},\ref{aff145}}
\and C.~C.~Kirkpatrick\inst{\ref{aff75}}
\and S.~Kruk\orcid{0000-0001-8010-8879}\inst{\ref{aff25}}
\and M.~Lattanzi\orcid{0000-0003-1059-2532}\inst{\ref{aff121}}
\and A.~M.~C.~Le~Brun\orcid{0000-0002-0936-4594}\inst{\ref{aff118}}
\and J.~Le~Graet\orcid{0000-0001-6523-7971}\inst{\ref{aff76}}
\and L.~Legrand\orcid{0000-0003-0610-5252}\inst{\ref{aff146},\ref{aff147}}
\and M.~Lembo\orcid{0000-0002-5271-5070}\inst{\ref{aff120},\ref{aff121}}
\and J.~Lesgourgues\orcid{0000-0001-7627-353X}\inst{\ref{aff45}}
\and T.~I.~Liaudat\orcid{0000-0002-9104-314X}\inst{\ref{aff148}}
\and S.~J.~Liu\orcid{0000-0001-7680-2139}\inst{\ref{aff59}}
\and A.~Loureiro\orcid{0000-0002-4371-0876}\inst{\ref{aff149},\ref{aff150}}
\and M.~Magliocchetti\orcid{0000-0001-9158-4838}\inst{\ref{aff59}}
\and F.~Mannucci\orcid{0000-0002-4803-2381}\inst{\ref{aff151}}
\and R.~Maoli\orcid{0000-0002-6065-3025}\inst{\ref{aff152},\ref{aff6}}
\and J.~Mart\'{i}n-Fleitas\orcid{0000-0002-8594-569X}\inst{\ref{aff111}}
\and C.~J.~A.~P.~Martins\orcid{0000-0002-4886-9261}\inst{\ref{aff153},\ref{aff154}}
\and L.~Maurin\orcid{0000-0002-8406-0857}\inst{\ref{aff24}}
\and R.~B.~Metcalf\orcid{0000-0003-3167-2574}\inst{\ref{aff19},\ref{aff14}}
\and M.~Miluzio\inst{\ref{aff25},\ref{aff155}}
\and P.~Monaco\orcid{0000-0003-2083-7564}\inst{\ref{aff125},\ref{aff17},\ref{aff28},\ref{aff18}}
\and C.~Moretti\orcid{0000-0003-3314-8936}\inst{\ref{aff29},\ref{aff113},\ref{aff17},\ref{aff18},\ref{aff28}}
\and G.~Morgante\inst{\ref{aff14}}
\and C.~Murray\inst{\ref{aff87}}
\and K.~Naidoo\orcid{0000-0002-9182-1802}\inst{\ref{aff46}}
\and A.~Navarro-Alsina\orcid{0000-0002-3173-2592}\inst{\ref{aff85}}
\and S.~Nesseris\orcid{0000-0002-0567-0324}\inst{\ref{aff114}}
\and K.~Paterson\orcid{0000-0001-8340-3486}\inst{\ref{aff73}}
\and L.~Patrizii\inst{\ref{aff15}}
\and A.~Pisani\orcid{0000-0002-6146-4437}\inst{\ref{aff76},\ref{aff156}}
\and V.~Popa\orcid{0000-0002-9118-8330}\inst{\ref{aff93}}
\and D.~Potter\orcid{0000-0002-0757-5195}\inst{\ref{aff157}}
\and I.~Risso\orcid{0000-0003-2525-7761}\inst{\ref{aff158}}
\and P.-F.~Rocci\inst{\ref{aff24}}
\and M.~Sahl\'en\orcid{0000-0003-0973-4804}\inst{\ref{aff159}}
\and E.~Sarpa\orcid{0000-0002-1256-655X}\inst{\ref{aff29},\ref{aff113},\ref{aff28}}
\and A.~Schneider\orcid{0000-0001-7055-8104}\inst{\ref{aff157}}
\and M.~Schultheis\inst{\ref{aff16}}
\and D.~Sciotti\orcid{0009-0008-4519-2620}\inst{\ref{aff6},\ref{aff42}}
\and E.~Sellentin\inst{\ref{aff160},\ref{aff66}}
\and L.~C.~Smith\orcid{0000-0002-3259-2771}\inst{\ref{aff161}}
\and S.~A.~Stanford\orcid{0000-0003-0122-0841}\inst{\ref{aff162}}
\and K.~Tanidis\inst{\ref{aff110}}
\and C.~Tao\orcid{0000-0001-7961-8177}\inst{\ref{aff76}}
\and G.~Testera\inst{\ref{aff35}}
\and R.~Teyssier\orcid{0000-0001-7689-0933}\inst{\ref{aff156}}
\and S.~Toft\orcid{0000-0003-3631-7176}\inst{\ref{aff163},\ref{aff164}}
\and S.~Tosi\orcid{0000-0002-7275-9193}\inst{\ref{aff34},\ref{aff35}}
\and A.~Troja\orcid{0000-0003-0239-4595}\inst{\ref{aff94},\ref{aff60}}
\and M.~Tucci\inst{\ref{aff58}}
\and C.~Valieri\inst{\ref{aff15}}
\and J.~Valiviita\orcid{0000-0001-6225-3693}\inst{\ref{aff78},\ref{aff79}}
\and D.~Vergani\orcid{0000-0003-0898-2216}\inst{\ref{aff14}}
\and G.~Verza\orcid{0000-0002-1886-8348}\inst{\ref{aff165}}
\and P.~Vielzeuf\orcid{0000-0003-2035-9339}\inst{\ref{aff76}}
\and N.~A.~Walton\orcid{0000-0003-3983-8778}\inst{\ref{aff161}}}
										   
\institute{School of Physics and Astronomy, University of Nottingham, University Park, Nottingham NG7 2RD, UK\label{aff1}
\and
Max Planck Institute for Extraterrestrial Physics, Giessenbachstr. 1, 85748 Garching, Germany\label{aff2}
\and
OCA, P.H.C Boulevard de l'Observatoire CS 34229, 06304 Nice Cedex 4, France\label{aff3}
\and
Waterloo Centre for Astrophysics, University of Waterloo, Waterloo, Ontario N2L 3G1, Canada\label{aff4}
\and
Department of Physics and Astronomy, University of Waterloo, Waterloo, Ontario N2L 3G1, Canada\label{aff5}
\and
INAF-Osservatorio Astronomico di Roma, Via Frascati 33, 00078 Monteporzio Catone, Italy\label{aff6}
\and
Observatorio Nacional, Rua General Jose Cristino, 77-Bairro Imperial de Sao Cristovao, Rio de Janeiro, 20921-400, Brazil\label{aff7}
\and
Institut d'Astrophysique de Paris, 98bis Boulevard Arago, 75014, Paris, France\label{aff8}
\and
Department of Astronomy, University of Florida, Bryant Space Science Center, Gainesville, FL 32611, USA\label{aff9}
\and
Instituto de Astrof\'isica de Andaluc\'ia, CSIC, Glorieta de la Astronom\'\i a, 18080, Granada, Spain\label{aff10}
\and
Institute of Space Sciences (ICE, CSIC), Campus UAB, Carrer de Can Magrans, s/n, 08193 Barcelona, Spain\label{aff11}
\and
Instituto de Astrof\'isica de Canarias, Calle V\'ia L\'actea s/n, 38204, San Crist\'obal de La Laguna, Tenerife, Spain\label{aff12}
\and
Departamento de Astrof\'isica, Universidad de La Laguna, 38206, La Laguna, Tenerife, Spain\label{aff13}
\and
INAF-Osservatorio di Astrofisica e Scienza dello Spazio di Bologna, Via Piero Gobetti 93/3, 40129 Bologna, Italy\label{aff14}
\and
INFN-Sezione di Bologna, Viale Berti Pichat 6/2, 40127 Bologna, Italy\label{aff15}
\and
Universit\'e C\^{o}te d'Azur, Observatoire de la C\^{o}te d'Azur, CNRS, Laboratoire Lagrange, Bd de l'Observatoire, CS 34229, 06304 Nice cedex 4, France\label{aff16}
\and
INAF-Osservatorio Astronomico di Trieste, Via G. B. Tiepolo 11, 34143 Trieste, Italy\label{aff17}
\and
IFPU, Institute for Fundamental Physics of the Universe, via Beirut 2, 34151 Trieste, Italy\label{aff18}
\and
Dipartimento di Fisica e Astronomia "Augusto Righi" - Alma Mater Studiorum Universit\`a di Bologna, via Piero Gobetti 93/2, 40129 Bologna, Italy\label{aff19}
\and
Aix-Marseille Universit\'e, CNRS, CNES, LAM, Marseille, France\label{aff20}
\and
Universit\'e de Strasbourg, CNRS, Observatoire astronomique de Strasbourg, UMR 7550, 67000 Strasbourg, France\label{aff21}
\and
NASA Ames Research Center, Moffett Field, CA 94035, USA\label{aff22}
\and
Bay Area Environmental Research Institute, Moffett Field, California 94035, USA\label{aff23}
\and
Universit\'e Paris-Saclay, CNRS, Institut d'astrophysique spatiale, 91405, Orsay, France\label{aff24}
\and
ESAC/ESA, Camino Bajo del Castillo, s/n., Urb. Villafranca del Castillo, 28692 Villanueva de la Ca\~nada, Madrid, Spain\label{aff25}
\and
School of Mathematics and Physics, University of Surrey, Guildford, Surrey, GU2 7XH, UK\label{aff26}
\and
INAF-Osservatorio Astronomico di Brera, Via Brera 28, 20122 Milano, Italy\label{aff27}
\and
INFN, Sezione di Trieste, Via Valerio 2, 34127 Trieste TS, Italy\label{aff28}
\and
SISSA, International School for Advanced Studies, Via Bonomea 265, 34136 Trieste TS, Italy\label{aff29}
\and
Dipartimento di Fisica e Astronomia, Universit\`a di Bologna, Via Gobetti 93/2, 40129 Bologna, Italy\label{aff30}
\and
Centre National d'Etudes Spatiales -- Centre spatial de Toulouse, 18 avenue Edouard Belin, 31401 Toulouse Cedex 9, France\label{aff31}
\and
Universit\"ats-Sternwarte M\"unchen, Fakult\"at f\"ur Physik, Ludwig-Maximilians-Universit\"at M\"unchen, Scheinerstrasse 1, 81679 M\"unchen, Germany\label{aff32}
\and
INAF-Osservatorio Astrofisico di Torino, Via Osservatorio 20, 10025 Pino Torinese (TO), Italy\label{aff33}
\and
Dipartimento di Fisica, Universit\`a di Genova, Via Dodecaneso 33, 16146, Genova, Italy\label{aff34}
\and
INFN-Sezione di Genova, Via Dodecaneso 33, 16146, Genova, Italy\label{aff35}
\and
Department of Physics "E. Pancini", University Federico II, Via Cinthia 6, 80126, Napoli, Italy\label{aff36}
\and
INAF-Osservatorio Astronomico di Capodimonte, Via Moiariello 16, 80131 Napoli, Italy\label{aff37}
\and
INFN section of Naples, Via Cinthia 6, 80126, Napoli, Italy\label{aff38}
\and
Dipartimento di Fisica, Universit\`a degli Studi di Torino, Via P. Giuria 1, 10125 Torino, Italy\label{aff39}
\and
INFN-Sezione di Torino, Via P. Giuria 1, 10125 Torino, Italy\label{aff40}
\and
INAF-IASF Milano, Via Alfonso Corti 12, 20133 Milano, Italy\label{aff41}
\and
INFN-Sezione di Roma, Piazzale Aldo Moro, 2 - c/o Dipartimento di Fisica, Edificio G. Marconi, 00185 Roma, Italy\label{aff42}
\and
Centro de Investigaciones Energ\'eticas, Medioambientales y Tecnol\'ogicas (CIEMAT), Avenida Complutense 40, 28040 Madrid, Spain\label{aff43}
\and
Port d'Informaci\'{o} Cient\'{i}fica, Campus UAB, C. Albareda s/n, 08193 Bellaterra (Barcelona), Spain\label{aff44}
\and
Institute for Theoretical Particle Physics and Cosmology (TTK), RWTH Aachen University, 52056 Aachen, Germany\label{aff45}
\and
Institute of Cosmology and Gravitation, University of Portsmouth, Portsmouth PO1 3FX, UK\label{aff46}
\and
Dipartimento di Fisica e Astronomia "Augusto Righi" - Alma Mater Studiorum Universit\`a di Bologna, Viale Berti Pichat 6/2, 40127 Bologna, Italy\label{aff47}
\and
Institute for Astronomy, University of Edinburgh, Royal Observatory, Blackford Hill, Edinburgh EH9 3HJ, UK\label{aff48}
\and
Jodrell Bank Centre for Astrophysics, Department of Physics and Astronomy, University of Manchester, Oxford Road, Manchester M13 9PL, UK\label{aff49}
\and
European Space Agency/ESRIN, Largo Galileo Galilei 1, 00044 Frascati, Roma, Italy\label{aff50}
\and
Universit\'e Claude Bernard Lyon 1, CNRS/IN2P3, IP2I Lyon, UMR 5822, Villeurbanne, F-69100, France\label{aff51}
\and
Institut de Ci\`{e}ncies del Cosmos (ICCUB), Universitat de Barcelona (IEEC-UB), Mart\'{i} i Franqu\`{e}s 1, 08028 Barcelona, Spain\label{aff52}
\and
Instituci\'o Catalana de Recerca i Estudis Avan\c{c}ats (ICREA), Passeig de Llu\'{\i}s Companys 23, 08010 Barcelona, Spain\label{aff53}
\and
UCB Lyon 1, CNRS/IN2P3, IUF, IP2I Lyon, 4 rue Enrico Fermi, 69622 Villeurbanne, France\label{aff54}
\and
Universit\'e Paris-Saclay, Universit\'e Paris Cit\'e, CEA, CNRS, AIM, 91191, Gif-sur-Yvette, France\label{aff55}
\and
Departamento de F\'isica, Faculdade de Ci\^encias, Universidade de Lisboa, Edif\'icio C8, Campo Grande, PT1749-016 Lisboa, Portugal\label{aff56}
\and
Instituto de Astrof\'isica e Ci\^encias do Espa\c{c}o, Faculdade de Ci\^encias, Universidade de Lisboa, Campo Grande, 1749-016 Lisboa, Portugal\label{aff57}
\and
Department of Astronomy, University of Geneva, ch. d'Ecogia 16, 1290 Versoix, Switzerland\label{aff58}
\and
INAF-Istituto di Astrofisica e Planetologia Spaziali, via del Fosso del Cavaliere, 100, 00100 Roma, Italy\label{aff59}
\and
INFN-Padova, Via Marzolo 8, 35131 Padova, Italy\label{aff60}
\and
Space Science Data Center, Italian Space Agency, via del Politecnico snc, 00133 Roma, Italy\label{aff61}
\and
School of Physics, HH Wills Physics Laboratory, University of Bristol, Tyndall Avenue, Bristol, BS8 1TL, UK\label{aff62}
\and
FRACTAL S.L.N.E., calle Tulip\'an 2, Portal 13 1A, 28231, Las Rozas de Madrid, Spain\label{aff63}
\and
INAF-Osservatorio Astronomico di Padova, Via dell'Osservatorio 5, 35122 Padova, Italy\label{aff64}
\and
Institute of Theoretical Astrophysics, University of Oslo, P.O. Box 1029 Blindern, 0315 Oslo, Norway\label{aff65}
\and
Leiden Observatory, Leiden University, Einsteinweg 55, 2333 CC Leiden, The Netherlands\label{aff66}
\and
Jet Propulsion Laboratory, California Institute of Technology, 4800 Oak Grove Drive, Pasadena, CA, 91109, USA\label{aff67}
\and
Department of Physics, Lancaster University, Lancaster, LA1 4YB, UK\label{aff68}
\and
Felix Hormuth Engineering, Goethestr. 17, 69181 Leimen, Germany\label{aff69}
\and
Technical University of Denmark, Elektrovej 327, 2800 Kgs. Lyngby, Denmark\label{aff70}
\and
Cosmic Dawn Center (DAWN), Denmark\label{aff71}
\and
Institut d'Astrophysique de Paris, UMR 7095, CNRS, and Sorbonne Universit\'e, 98 bis boulevard Arago, 75014 Paris, France\label{aff72}
\and
Max-Planck-Institut f\"ur Astronomie, K\"onigstuhl 17, 69117 Heidelberg, Germany\label{aff73}
\and
NASA Goddard Space Flight Center, Greenbelt, MD 20771, USA\label{aff74}
\and
Department of Physics and Helsinki Institute of Physics, Gustaf H\"allstr\"omin katu 2, 00014 University of Helsinki, Finland\label{aff75}
\and
Aix-Marseille Universit\'e, CNRS/IN2P3, CPPM, Marseille, France\label{aff76}
\and
Universit\'e de Gen\`eve, D\'epartement de Physique Th\'eorique and Centre for Astroparticle Physics, 24 quai Ernest-Ansermet, CH-1211 Gen\`eve 4, Switzerland\label{aff77}
\and
Department of Physics, P.O. Box 64, 00014 University of Helsinki, Finland\label{aff78}
\and
Helsinki Institute of Physics, Gustaf H{\"a}llstr{\"o}min katu 2, University of Helsinki, Helsinki, Finland\label{aff79}
\and
Mullard Space Science Laboratory, University College London, Holmbury St Mary, Dorking, Surrey RH5 6NT, UK\label{aff80}
\and
NOVA optical infrared instrumentation group at ASTRON, Oude Hoogeveensedijk 4, 7991PD, Dwingeloo, The Netherlands\label{aff81}
\and
Centre de Calcul de l'IN2P3/CNRS, 21 avenue Pierre de Coubertin 69627 Villeurbanne Cedex, France\label{aff82}
\and
Dipartimento di Fisica "Aldo Pontremoli", Universit\`a degli Studi di Milano, Via Celoria 16, 20133 Milano, Italy\label{aff83}
\and
INFN-Sezione di Milano, Via Celoria 16, 20133 Milano, Italy\label{aff84}
\and
Universit\"at Bonn, Argelander-Institut f\"ur Astronomie, Auf dem H\"ugel 71, 53121 Bonn, Germany\label{aff85}
\and
Department of Physics, Institute for Computational Cosmology, Durham University, South Road, Durham, DH1 3LE, UK\label{aff86}
\and
Universit\'e Paris Cit\'e, CNRS, Astroparticule et Cosmologie, 75013 Paris, France\label{aff87}
\and
University of Applied Sciences and Arts of Northwestern Switzerland, School of Engineering, 5210 Windisch, Switzerland\label{aff88}
\and
Institute of Physics, Laboratory of Astrophysics, Ecole Polytechnique F\'ed\'erale de Lausanne (EPFL), Observatoire de Sauverny, 1290 Versoix, Switzerland\label{aff89}
\and
Institut de F\'{i}sica d'Altes Energies (IFAE), The Barcelona Institute of Science and Technology, Campus UAB, 08193 Bellaterra (Barcelona), Spain\label{aff90}
\and
European Space Agency/ESTEC, Keplerlaan 1, 2201 AZ Noordwijk, The Netherlands\label{aff91}
\and
DARK, Niels Bohr Institute, University of Copenhagen, Jagtvej 155, 2200 Copenhagen, Denmark\label{aff92}
\and
Institute of Space Science, Str. Atomistilor, nr. 409 M\u{a}gurele, Ilfov, 077125, Romania\label{aff93}
\and
Dipartimento di Fisica e Astronomia "G. Galilei", Universit\`a di Padova, Via Marzolo 8, 35131 Padova, Italy\label{aff94}
\and
Institut f\"ur Theoretische Physik, University of Heidelberg, Philosophenweg 16, 69120 Heidelberg, Germany\label{aff95}
\and
Institut de Recherche en Astrophysique et Plan\'etologie (IRAP), Universit\'e de Toulouse, CNRS, UPS, CNES, 14 Av. Edouard Belin, 31400 Toulouse, France\label{aff96}
\and
Universit\'e St Joseph; Faculty of Sciences, Beirut, Lebanon\label{aff97}
\and
Departamento de F\'isica, FCFM, Universidad de Chile, Blanco Encalada 2008, Santiago, Chile\label{aff98}
\and
Universit\"at Innsbruck, Institut f\"ur Astro- und Teilchenphysik, Technikerstr. 25/8, 6020 Innsbruck, Austria\label{aff99}
\and
Institut d'Estudis Espacials de Catalunya (IEEC),  Edifici RDIT, Campus UPC, 08860 Castelldefels, Barcelona, Spain\label{aff100}
\and
Satlantis, University Science Park, Sede Bld 48940, Leioa-Bilbao, Spain\label{aff101}
\and
Instituto de Astrof\'isica e Ci\^encias do Espa\c{c}o, Faculdade de Ci\^encias, Universidade de Lisboa, Tapada da Ajuda, 1349-018 Lisboa, Portugal\label{aff102}
\and
Universidad Polit\'ecnica de Cartagena, Departamento de Electr\'onica y Tecnolog\'ia de Computadoras,  Plaza del Hospital 1, 30202 Cartagena, Spain\label{aff103}
\and
Centre for Information Technology, University of Groningen, P.O. Box 11044, 9700 CA Groningen, The Netherlands\label{aff104}
\and
INFN-Bologna, Via Irnerio 46, 40126 Bologna, Italy\label{aff105}
\and
Kapteyn Astronomical Institute, University of Groningen, PO Box 800, 9700 AV Groningen, The Netherlands\label{aff106}
\and
Infrared Processing and Analysis Center, California Institute of Technology, Pasadena, CA 91125, USA\label{aff107}
\and
INAF, Istituto di Radioastronomia, Via Piero Gobetti 101, 40129 Bologna, Italy\label{aff108}
\and
Astronomical Observatory of the Autonomous Region of the Aosta Valley (OAVdA), Loc. Lignan 39, I-11020, Nus (Aosta Valley), Italy\label{aff109}
\and
Department of Physics, Oxford University, Keble Road, Oxford OX1 3RH, UK\label{aff110}
\and
Aurora Technology for European Space Agency (ESA), Camino bajo del Castillo, s/n, Urbanizacion Villafranca del Castillo, Villanueva de la Ca\~nada, 28692 Madrid, Spain\label{aff111}
\and
ICL, Junia, Universit\'e Catholique de Lille, LITL, 59000 Lille, France\label{aff112}
\and
ICSC - Centro Nazionale di Ricerca in High Performance Computing, Big Data e Quantum Computing, Via Magnanelli 2, Bologna, Italy\label{aff113}
\and
Instituto de F\'isica Te\'orica UAM-CSIC, Campus de Cantoblanco, 28049 Madrid, Spain\label{aff114}
\and
CERCA/ISO, Department of Physics, Case Western Reserve University, 10900 Euclid Avenue, Cleveland, OH 44106, USA\label{aff115}
\and
Technical University of Munich, TUM School of Natural Sciences, Physics Department, James-Franck-Str.~1, 85748 Garching, Germany\label{aff116}
\and
Max-Planck-Institut f\"ur Astrophysik, Karl-Schwarzschild-Str.~1, 85748 Garching, Germany\label{aff117}
\and
Laboratoire Univers et Th\'eorie, Observatoire de Paris, Universit\'e PSL, Universit\'e Paris Cit\'e, CNRS, 92190 Meudon, France\label{aff118}
\and
Departamento de F{\'\i}sica Fundamental. Universidad de Salamanca. Plaza de la Merced s/n. 37008 Salamanca, Spain\label{aff119}
\and
Dipartimento di Fisica e Scienze della Terra, Universit\`a degli Studi di Ferrara, Via Giuseppe Saragat 1, 44122 Ferrara, Italy\label{aff120}
\and
Istituto Nazionale di Fisica Nucleare, Sezione di Ferrara, Via Giuseppe Saragat 1, 44122 Ferrara, Italy\label{aff121}
\and
Center for Data-Driven Discovery, Kavli IPMU (WPI), UTIAS, The University of Tokyo, Kashiwa, Chiba 277-8583, Japan\label{aff122}
\and
Ludwig-Maximilians-University, Schellingstrasse 4, 80799 Munich, Germany\label{aff123}
\and
Max-Planck-Institut f\"ur Physik, Boltzmannstr. 8, 85748 Garching, Germany\label{aff124}
\and
Dipartimento di Fisica - Sezione di Astronomia, Universit\`a di Trieste, Via Tiepolo 11, 34131 Trieste, Italy\label{aff125}
\and
California institute of Technology, 1200 E California Blvd, Pasadena, CA 91125, USA\label{aff126}
\and
Institute Lorentz, Leiden University, Niels Bohrweg 2, 2333 CA Leiden, The Netherlands\label{aff127}
\and
Institute for Astronomy, University of Hawaii, 2680 Woodlawn Drive, Honolulu, HI 96822, USA\label{aff128}
\and
Department of Physics \& Astronomy, University of California Irvine, Irvine CA 92697, USA\label{aff129}
\and
Department of Mathematics and Physics E. De Giorgi, University of Salento, Via per Arnesano, CP-I93, 73100, Lecce, Italy\label{aff130}
\and
INFN, Sezione di Lecce, Via per Arnesano, CP-193, 73100, Lecce, Italy\label{aff131}
\and
INAF-Sezione di Lecce, c/o Dipartimento Matematica e Fisica, Via per Arnesano, 73100, Lecce, Italy\label{aff132}
\and
Departamento F\'isica Aplicada, Universidad Polit\'ecnica de Cartagena, Campus Muralla del Mar, 30202 Cartagena, Murcia, Spain\label{aff133}
\and
Instituto de Astrof\'isica de Canarias (IAC); Departamento de Astrof\'isica, Universidad de La Laguna (ULL), 38200, La Laguna, Tenerife, Spain\label{aff134}
\and
Instituto de F\'isica de Cantabria, Edificio Juan Jord\'a, Avenida de los Castros, 39005 Santander, Spain\label{aff135}
\and
Department of Computer Science, Aalto University, PO Box 15400, Espoo, FI-00 076, Finland\label{aff136}
\and
Instituto de Astrof\'\i sica de Canarias, c/ Via Lactea s/n, La Laguna E-38200, Spain. Departamento de Astrof\'\i sica de la Universidad de La Laguna, Avda. Francisco Sanchez, La Laguna, E-38200, Spain\label{aff137}
\and
Caltech/IPAC, 1200 E. California Blvd., Pasadena, CA 91125, USA\label{aff138}
\and
Ruhr University Bochum, Faculty of Physics and Astronomy, Astronomical Institute (AIRUB), German Centre for Cosmological Lensing (GCCL), 44780 Bochum, Germany\label{aff139}
\and
Univ. Grenoble Alpes, CNRS, Grenoble INP, LPSC-IN2P3, 53, Avenue des Martyrs, 38000, Grenoble, France\label{aff140}
\and
Department of Physics and Astronomy, Vesilinnantie 5, 20014 University of Turku, Finland\label{aff141}
\and
Serco for European Space Agency (ESA), Camino bajo del Castillo, s/n, Urbanizacion Villafranca del Castillo, Villanueva de la Ca\~nada, 28692 Madrid, Spain\label{aff142}
\and
ARC Centre of Excellence for Dark Matter Particle Physics, Melbourne, Australia\label{aff143}
\and
Centre for Astrophysics \& Supercomputing, Swinburne University of Technology,  Hawthorn, Victoria 3122, Australia\label{aff144}
\and
Department of Physics and Astronomy, University of the Western Cape, Bellville, Cape Town, 7535, South Africa\label{aff145}
\and
DAMTP, Centre for Mathematical Sciences, Wilberforce Road, Cambridge CB3 0WA, UK\label{aff146}
\and
Kavli Institute for Cosmology Cambridge, Madingley Road, Cambridge, CB3 0HA, UK\label{aff147}
\and
IRFU, CEA, Universit\'e Paris-Saclay 91191 Gif-sur-Yvette Cedex, France\label{aff148}
\and
Oskar Klein Centre for Cosmoparticle Physics, Department of Physics, Stockholm University, Stockholm, SE-106 91, Sweden\label{aff149}
\and
Astrophysics Group, Blackett Laboratory, Imperial College London, London SW7 2AZ, UK\label{aff150}
\and
INAF-Osservatorio Astrofisico di Arcetri, Largo E. Fermi 5, 50125, Firenze, Italy\label{aff151}
\and
Dipartimento di Fisica, Sapienza Universit\`a di Roma, Piazzale Aldo Moro 2, 00185 Roma, Italy\label{aff152}
\and
Centro de Astrof\'{\i}sica da Universidade do Porto, Rua das Estrelas, 4150-762 Porto, Portugal\label{aff153}
\and
Instituto de Astrof\'isica e Ci\^encias do Espa\c{c}o, Universidade do Porto, CAUP, Rua das Estrelas, PT4150-762 Porto, Portugal\label{aff154}
\and
HE Space for European Space Agency (ESA), Camino bajo del Castillo, s/n, Urbanizacion Villafranca del Castillo, Villanueva de la Ca\~nada, 28692 Madrid, Spain\label{aff155}
\and
Department of Astrophysical Sciences, Peyton Hall, Princeton University, Princeton, NJ 08544, USA\label{aff156}
\and
Department of Astrophysics, University of Zurich, Winterthurerstrasse 190, 8057 Zurich, Switzerland\label{aff157}
\and
INAF-Osservatorio Astronomico di Brera, Via Brera 28, 20122 Milano, Italy, and INFN-Sezione di Genova, Via Dodecaneso 33, 16146, Genova, Italy\label{aff158}
\and
Theoretical astrophysics, Department of Physics and Astronomy, Uppsala University, Box 515, 751 20 Uppsala, Sweden\label{aff159}
\and
Mathematical Institute, University of Leiden, Niels Bohrweg 1, 2333 CA Leiden, The Netherlands\label{aff160}
\and
Institute of Astronomy, University of Cambridge, Madingley Road, Cambridge CB3 0HA, UK\label{aff161}
\and
Department of Physics and Astronomy, University of California, Davis, CA 95616, USA\label{aff162}
\and
Cosmic Dawn Center (DAWN)\label{aff163}
\and
Niels Bohr Institute, University of Copenhagen, Jagtvej 128, 2200 Copenhagen, Denmark\label{aff164}
\and
Center for Computational Astrophysics, Flatiron Institute, 162 5th Avenue, 10010, New York, NY, USA\label{aff165}}


%
%
   \abstract{

The intracluster light (ICL) permeating galaxy clusters is a tracer of the cluster's assembly history, and potentially a tracer of their dark matter structure.
In this work we explore the capability of the Euclid Wide Survey to detect ICL using \eceb{\HE-band} mock images. We simulate clusters across a range of redshifts (0.3--1.8) and halo masses ($10^{13.9}$--$10^{15.0}\,\Msun$), using an observationally motivated model of the ICL. We identify a 50--200\,kpc circular annulus around the brightest cluster galaxy (BCG) in which the signal-to-noise ratio (S/N) of the ICL is maximised and use the S/N within this aperture as our figure of merit for ICL detection. We compare three state-of-the-art methods for ICL detection, and find that a method that performs simple aperture photometry after high-surface brightness source masking is able to detect ICL with minimal bias for clusters more massive than $10^{14.2}\,\Msun$.
The S/N of the ICL detection is primarily limited by the redshift of the cluster, driven by cosmological dimming, rather than the mass of the cluster. Assuming the ICL in each cluster contains 15\% of the stellar light, we forecast that \Euclid will be able to measure the presence of ICL in \eceb{up to} $\sim80$\,000 clusters of $>10^{14.2}\,\Msun$ between $z=0.3$ and 1.5 with a $\SN>3$. Half of these clusters will reside below $z=0.75$ and the majority of those below $z=0.6$ will be detected with a $\SN>20$. A few thousand clusters at $1.3<z<1.5$ will have ICL detectable with a S/N greater than 3. 
The surface brightness profile of the ICL model is strongly dependent on both the mass of the cluster and the redshift at which it is observed so the outer ICL is best observed in the most massive clusters of $>10^{14.7}\,\Msun$.  \Euclid will detect the ICL at more than 500\,kpc distance from the BCG, up to $z=0.7$, in several hundred of these massive clusters over its large survey volume. 
} 
%
%
%
   \titlerunning{Forecasting ICL detection limits in the EWS}
   \authorrunning{Euclid Collaboration: Bellhouse et al.}
   
   \maketitle
%
%
%
%

\section{\label{sc:Intro}Introduction}

Intracluster light (ICL) is a powerful diagnostic tracer of the assembly history of galaxy clusters that offers insight into the hierarchical formation of clusters and the dynamical evolution of their member galaxies \citep[e.g.,][]{gonzalez2000,feldmeier2004, mihos2005,krick2007,murante07, watkins2014, golden-marx23, golden-marx2024}. The ICL forms as a result of hierarchical mergers, as well as tidal interactions between member galaxies and the cluster's gravitational field. The formation of the brightest cluster galaxy (BCG) and the ICL is closely linked \citep{murante07, montes2022a, chun2023, golden-marx2024, contini2024}, both of which form through hierarchical accretion in similar ways \citep{Amorisco2017,pillepich2018}. The quantity, structure, and evolution of the ICL therefore provides a test of our models of galaxy and cluster formation.

The dominant component of galaxy clusters is dark matter, but this dark matter is challenging to detect observationally, with the most direct method being gravitational lensing \citep{massey2010,kneib2011,hoekstra2013}. In recent years however, studies have suggested that ICL can be used as a dark matter tracer. \citet{jee2010} suggests that the ICL traces the gravitational potential of the cluster in a manner similar to dark matter, while \citet{montes2019} found the scale radii and projected shape of the dark matter halo and ICL to be in agreement. Furthermore, \citet{Pulsoni2021b} used simulations to show that the ICL can be used to infer the intrinsic shape and principal axes of the dark matter halo. This correlation, \referee{further reinforced in more recent simulations \citep{Yoo2024,Contreras-Santos2024},} opens up the possibility of carrying out detailed studies of the dark matter distribution in clusters solely using deep photometric observations.

\eceb{ICL may be useful for confirming that optically-selected clusters are indeed collapsed massive halos.} Optically identifying a pure sample of galaxy clusters in large photometric surveys, particularly beyond $z > 0.3$, becomes challenging \eceb{due to projection effects \citep{Zu2017,Myles2021,Wu2022}, which can lead to a spurious overdensity of galaxies along a line-of-sight being mistakenly identified as a cluster}.
\referee{While X-ray imaging surveys \citep{fassbender2011,mehrtens2012,mantz2018} and those based on the Sunyaev--Zeldovich effect \citep{carlstrom2011,Hilton2021} can efficiently detect clusters at $z<1$, only a few dozen have been identified beyond this redshift \citep[e.g.][]{bleem2015,hennig2017,barrena2018,aguado-barahona2019,khullar2019,barrena2020,huang2020}. This scarcity might suggest either a low space density of high-redshift clusters or a lack of significant intra-cluster gas detectable by these methods.}
Since the formation of ICL is intrinsically tied to the growth of galaxy clusters, the presence of ICL can act as confirmation of a collapsed massive dark matter halo.
Thus, ICL observations are useful \referee{as a complementary method to SZ and X-ray detection} for determining the purity of photometrically-selected cluster catalogues.

\eceb{ICL can also be used to measure the edge or physical extent of clusters.} A physically motivated definition of a cluster's edge is the `splashback' radius, which is identified by a sharp break in the radial density profile of the dark matter halo \citep{diemer2014}. The splashback radius is the apocentre of material infalling into a cluster \citep{gunn1972, fillmore1984, bertschinger1985}, and is closely tied to cluster mass and accretion rate. Several observational studies have successfully measured the splashback radius for stacked clusters using galaxy surface density profiles \citep{bianconi2021} and weak gravitational lensing profiles \citep{Giocoli2024}. The discovery by \citet{gonzalez2021} of an inflection in the ICL profile of a high-mass cluster at a location similar to the expected splashback radius may be the first evidence that this parameter can be measured in individual clusters by means of deep observations of the ICL.

The low surface brightness nature of the ICL presents an observational challenge, necessitating observations that are both deep and wide to sufficiently resolve the diffuse emission and measure the background light well beyond the cluster's edge.
Although some observations of individual clusters have detected ICL at large radii~\citep{gonzalez2021}, most surveys \citep[e.g.~][]{kluge20, golden-marx23, golden-marx2024} have been able to measure the light profile of the ICL in individual clusters only up to a few hundred kpc from the BCG.  These measurements have been enhanced by stacking the ICL signal of clusters with similar halo masses to extend the ICL out to Mpc scales \citep{zibetti05,zhang19,santos21,chen22,zhang2024}.  However, \Euclid's photometric stability, well-controlled point source function (PSF), and suppression of scattered light may allow us to extend the radial range for individual observations of BCG+ICL profiles for large samples of clusters, thereby enhancing our ability to understand cluster accretion histories and the connection between ICL and dark matter.

\Euclid \citep{laureijs2011, EuclidSkyOverview} will image almost a third of the sky  across four photometric bands from the visible (\IE) to the near-infrared (\YE, \JE, \HE). \Euclid's faint detection limit \citep{Scaramella-EP1, EROData, ERONearbyGals}, makes it ideally suited to image the low surface brightness universe \citep{Borlaff-EP16}. In combination with the wide-field imaging design of the telescope, this will bring forth unprecedented levels of detail in future studies of the ICL, as demonstrated in \citet{EROPerseusICL}.

The focus of this paper will be the ${\sim}\,14\,000\,\mathrm{deg}^2$ Euclid Wide Survey \citep[EWS;][]{Scaramella-EP1} and the large sample of clusters we will be able to study with this data. \eceb {In addition to the EWS, \Euclid will observe three deep fields totalling 53 deg$^2$, which will be 6 times deeper than the EWS. The deeper data will allow us to explore a small sample of clusters to increased limiting radii, higher redshift and lower limiting ICL fractions, however the area covered by the combined deep fields will be significantly reduced compared to the wide survey.}

In this study, we forecast the halo mass and redshift limits to which the ICL can be explored with \Euclid in the EWS.
We restrict the scope of this paper to the \HE\ band with the NISP instrument only  \citep{EuclidSkyNISP}, since light at this wavelength is a better tracer of stellar mass compared to the other wavelengths measured by \Euclid. \eceb{Whilst the \IE\ band offers higher spatial resolution and a lower limiting surface brightness, the effect of Galactic cirrus is higher \citep{roman2020}, and the cirrus features appear more filamentary at the wavelength of \IE, rendering them more challenging to correct.}

\eceb{This analysis focuses on the photometric measurements of the ICL, and we treat the BCG and ICL as a single component.} The stellar population belonging to the ICL can be separated from the BCG kinematically \citep[][]{Dolag2010,Longobardi2013,Longobardi2015,Remus2017,Hartke2022}. However, these kinetically distinct stellar populations overlap in space so it is impossible to separate the BCG and ICL stellar populations using photometry alone. Nevertheless, it is possible to define regions in space in which the majority of the stellar population belong to either the BCG or ICL. Several methods to separate the BCG and ICL photometrically have been tested in \citet{brough24}, and they conclude that the stellar population beyond 100\,kpc from the BCG core is typically dominated by the ICL.
Due to the challenge of separating the BCG and ICL and the uncertainties introduced by attempting to do so,  we have chosen to forecast the detectability of the total light from the combined BCG and ICL, and refer to the combined signal as BCG+ICL, though our interest is in the detectability of the ICL component, which dominates at larger radii.

This paper is arranged as follows:
Section\,\ref{sec:mocks} describes the creation of mock images of clusters, BCGs, and ICL as observed by \Euclid. 
Although we do not separate the BCG signal from the ICL signal in this work, we wish to determine the detectability of the low surface brightness ICL, rather than the high surface brightness BCG. Therefore, in Sect.\,\ref{sec:SNProfiles} we explore the radial profiles of the ICL signal-to-noise ratio (S/N) and select an aperture in which the S/N of the ICL is maximised.
Section~\ref{sec:sngrids} presents our forecasting of the BCG+ICL S/N in clusters across the parameter space of halo mass, redshift and ICL fraction, as they would be detected in the EWS.
In Sect.\,\ref{sec:fitcomparison}, we measure the BCG+ICL flux in a subset of our mock cluster images using the ICL detection techniques of \citet{Kluge2023}, \citet{Ellien2021}, and \citet{ahad2023}, then determine the S/N limit to which current measurement techniques will reliably detect ICL in \Euclid \HE images.
To estimate the number of clusters for which we will be able to measure ICL with \Euclid, we combine our results with the cluster halo mass function in Sect.\,\ref{sec:haloestimates}. 
Section~\ref{sec:extent} explores the radial detection limit of the ICL within individual clusters.
Finally, Sect.\,\ref{sec:discussion} discusses the implications of our results for studying ICL with \Euclid and we outline the conclusions of this analysis in Sect.\,\ref{sec:conclusions}.
Throughout this paper we use a standard $\Lambda$CDM flat cosmology with $\Omega_\mathrm{m}=0.3$ and $H_0=70\,\mathrm{km}\,\mathrm{s}^{-1}\,\mathrm{Mpc}^{-1}$.

\section{Mock observations}\label{sec:mocks}
\subsection{Simulating clusters and ICL in \Euclid mock images}
\label{sec:clustersims}

\begin{figure*}[tbp]
    \centering
    \includegraphics[width=0.8\hsize]{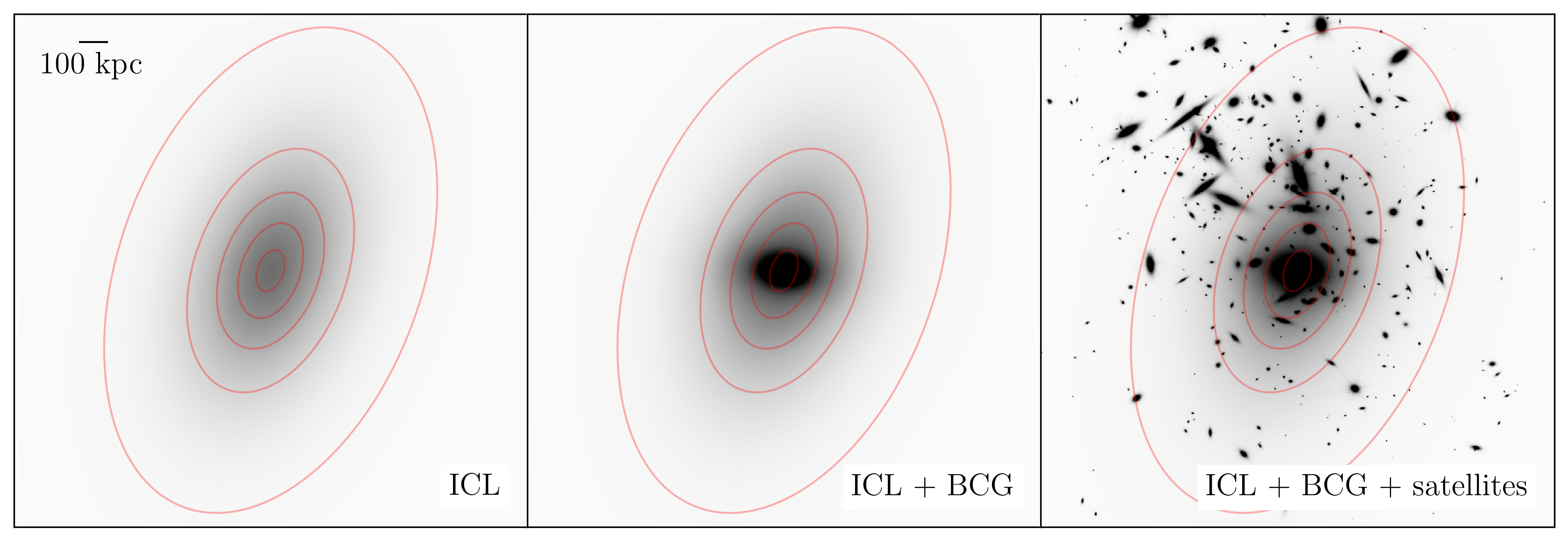}
    \caption{Components of the simulated cluster: the ICL only (left, and shown as contours in the middle and right panels), ICL combined with the BCG (middle), and ICL+BCG+satellite galaxies (right). The width of each square frame is 400\arcsec. The cluster has a halo mass of $10^{14.7}\,\Msun$ and is located at $z=0.3$. The ICL is simulated with $\fICL=0.15$, $n=0.76$ and $r_\mathrm{e}=300$\,kpc.}
    \label{fig:mock_components}
\end{figure*}
\begin{figure*}[tbp]
    \centering
    \includegraphics[width=0.8\hsize]{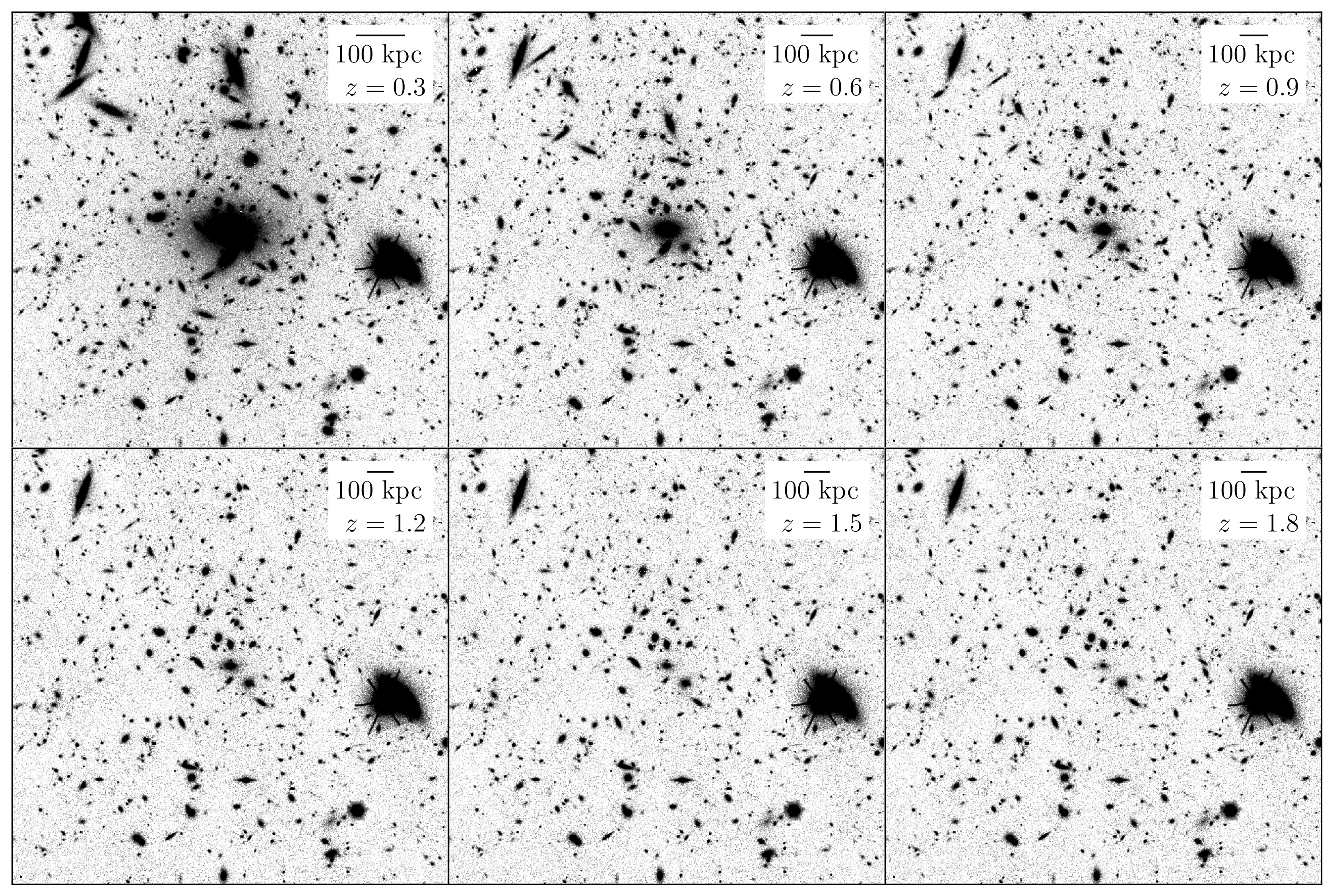}
    \caption{Examples of the mock EWS \HE\ mosaics centred on a simulated cluster with a halo mass of $10^{14.7}\Msun$, $\fICL=0.15$, $n=0.76$ and $r_\mathrm{e}=300$\,kpc. Each panel shows the exact same cluster (i.e.~the galaxies have not been evolved according to any star-formation history) rescaled and K-corrected to  the redshift in the legend. The width of each square frame is 200\arcsec.}
    \label{fig:mock_redshifts}
\end{figure*}

In order to forecast \Euclid's capability for detecting ICL we generated mock \Euclid\ observations of a sample of galaxy clusters which have a range of halo masses, $13.9 < \logMhalo < 15.0$, and redshifts ($0.3 < z < 1.8$), with various ICL properties, including the ICL fraction, S\'ersic index, and effective radii.

We selected sample clusters from the MAMBO (Mocks with Abundance Matching in Bologna) lightcone \citep{girelli2021} which has an area of $3.14$\,deg$^2$ and contains 101 clusters ($\Mhalo/\Msun \ge 10^{13.9}$). The cluster's halo mass, $\Mhalo$, is defined as the mass within $R_{200}$, the radius within which the average density is 200 times the critical density at the redshift of the cluster in the MAMBO catalogue, which approximately corresponds to the virial radius.
The MAMBO catalogue uses the halo and subhalo distribution from the Millennium simulation \citep{springel2005}, rescaling the halo properties to the \textit{Planck} cosmology using the method described in \citet{angulo2010}. Galaxy properties, such as morphology, size, and observed-frame photometry, are assigned to the subhalo and halo distribution via empirical prescriptions using the Empirical Galaxy Generator \citep[EGG;][]{Schreiber2017}. In order to maintain a uniform sample and minimise differences in evolutionary stages between clusters, we chose a sample of five clusters limited to a redshift range of $z = 0.51$--$0.77$. \eceb{Our focus of this analysis is on the variation of ICL properties rather than the differences between clusters, therefore we keep the cluster properties similar and concentrate on how ICL properties determine the detectability of ICL.}

To produce the mock images we made use of \texttt{galsim} \citep{rowe2015}. We selected satellite galaxies from each cluster defined in the MAMBO catalogue as all galaxies with the same halo ID except for \texttt{type=0} (defined in the catalogue as the central galaxy). For each cluster satellite galaxy, we superimposed two S\'ersic profiles with fixed S\'ersic indices of 3.5 and 1.5 for the bulge and disk respectively, taking the half-light radii for the bulge and disk as well as the bulge-to-disk ratio from the MAMBO catalogue. The position angle for each galaxy was randomly assigned. 

The central galaxies of massive clusters are a rare and special class of galaxy \citep{vonderLinden2007}. The prescription used to assign galaxy properties in the MAMBO catalogue is therefore unlikely to result in realistic BCG images. Furthermore, we wanted to ensure the close relationship between BCG and ICL properties was present in our mock images.  We therefore replaced the original central galaxy (object \texttt{type = 0}) from the MAMBO catalogue with an empirically defined BCG model constructed using the observational results presented in table 4 of \citet{kluge20}. Both the new BCG and the ICL components were drawn as S\'ersic profiles. The effective radius and S\'ersic index of the BCG were fixed to be $r_\mathrm{e,BCG}=22.6~\mathrm{kpc}$ and $n_\mathrm{BCG}=4.6$, respectively. The BCG flux was set at 15\% of the total cluster luminosity from stars at the wavelength of interest, \eceb{consistent with measurements in the literature \citep{kluge21}.} The axis ratio of the BCG was fixed at 0.6, within the range of the values measured by \citet{gonzalez2005}.

\eceb{We then added ICL to these mock clusters, simulating the ICL with an elliptical Sersic profile.} The effective radius of the ICL was allowed to vary such that it scaled with the total stellar mass of the cluster. We took the median value of the ICL effective radius ($r_{\mathrm{e},0}=189.5~\mathrm{kpc}$) from \citet{kluge20}, and scaled it using the stellar mass of the cluster according to a linear scaling relation (Eq.\,\ref{eq:scaling_re}) fit to the stellar masses and ICL effective radii of the sample of clusters from the same study,
\begin{equation}\label{eq:scaling_re}
\log_{10}\left(\frac{r_{\mathrm{e},\mathrm{scaled}}}{\mathrm{kpc}}\right) = [0.48\logMstar - 4.91] \log_{10}\left(\frac{r_{\mathrm{e},0}}{\mathrm{kpc}}\right) \,.
\end{equation}
The S\'ersic index of the ICL  was allowed to vary between $0.38<n_\mathrm{ICL}<1.52$, with the fiducial value at $n_\mathrm{ICL}=0.76$. \referee{The ranges and fiducial values of $n$ and $r$ are consistent with ranges of measured double-Sersic and Sersic-exponential fits of BCGs and ICL in the literature. \citet{Seigar2007} measured values of ($0.8 \lesssim n \lesssim 4$) and ($72.3~\mathrm{kpc} \lesssim r \lesssim 230~\mathrm{kpc}$) in a sample of five clusters at low redshift. At higher redshifts, \citet{joo2023} measured similar Sersic indices of ($0.7 \lesssim n \lesssim 3.7$) but lower scale radii ($15~\mathrm{kpc} \lesssim r \lesssim 75~\mathrm{kpc}$) across ten clusters at ($1 \lesssim z \lesssim 2$).}

The minor-to-major axis ratio of the ICL was set to 0.6, consistent with the average ICL ellipticities measured in \citet{kluge21} and the dark matter halo ellipticities measured in \citet{shin2018}.  We opted to randomly vary the position angle of the BCG with respect to the ICL as the literature shows the alignment between the components can vary significantly; \citet{gonzalez2005} \eceb{used two-component fitting to model the BCG and ICL in a sample of 24 low-redshift clusters and found only 40\% had an alignment between the BCG and the ICL, with the remainder exhibiting a strong misalignment.} The luminosity of the ICL was selected to contribute a given fraction of the total cluster luminosity ($\fICL$, a parameter which could be altered).
 These profiles of the BCG and ICL are best estimates based on the low-redshift sample of \citet{kluge20}, and it is not known how representative these will be for the true sample, as \Euclid will be the first mission that can fully explore \eceb{the near-infrared ICL for the range of redshifts, halo masses and ICL fractions considered in this analysis}.   

All the S\'ersic profiles were convolved with a model \Euclid PSF and sampled at the appropriate pixel scale and zeropoint for the \HE\ filter, resulting in a mock image of the isolated cluster at single-exposure depth. An example of the simulated ICL, BCG and the satellite galaxies constructed from a cluster in the MAMBO lightcone is shown in Fig.\,\ref{fig:mock_components}. In addition to the simulated sources, we also simulate the Poisson noise associated with these sources. This noise is added to the simulated sources (not shown in Fig.\,\ref{fig:mock_components}), and kept as a separate noise image for each exposure.

The simulated images of the BCG, ICL and satellite galaxies were then injected into mock \Euclid images of individual \HE\ exposures produced by \citet{EP-Serrano} for Science Challenge 8 (SC8). The SC8 mock \Euclid images include simulated galaxies from the Euclid Flagship Mock Galaxy catalogues (which include gravitational lensing effects), a realistic distribution of stars using the Besançon model \citep{robin1986}, and cosmic rays that occur during integration and detector readout. The modelled astronomical sources are combined with large-scale nuisance sources such as zodiacal light, thermal irradiance from spacecraft elements, diffuse scattered light, in- and out-of-field stray light, and optical reflections (ghosts).

Each mock exposure incorporates the filter transmission curves, geometric distortions, PSF, quantum efficiency, dark current, readout noise, and Poisson noise. Two sources that are important for ICL modelling are not included: charge persistence from the NISP detectors and Galactic cirri. \eceb{The impact of charge persistence is greatly reduced by subtraction, as shown in \citet{EROData}, and further masking. We discuss the impact of cirri in Sect.\,\ref{sec:haloestimates}. }

The simulated clusters were added to all four of the SC8 individual \HE\ exposures within a mock reference observation sequence (ROS) that simulates the sequence that will be used to observe the EWS. In this sequence, an image in \IE\ is taken simultaneously with slitless grism spectra by the NISP instrument, followed by NIR images taken in \JE, then \HE, and finally \YE\ all at the same telescope pointing. The telescope is then dithered slightly and the full observing sequence is repeated before the telescope dithers again. In total, four exposures of each field are taken through each filter and grism before the telescope is shifted to another field. We therefore add the simulated cluster to exactly the same world coordinates on each data frame of each exposure, while summing in quadrature the simulated noise exposures with the noise frames of each exposure.

We combined all four \HE\ exposures from a ROS using \texttt{SWARP} \citep{bertin2002} to produce a mosaic, replicating the equivalent step in the \Euclid data processing pipeline. \eceb{For ICL measurements, the optimal data processing constructs an image with the smoothest background possible, then removes a background measured on a scale that is large enough that it does not remove any ICL. The \Euclid\ data can be processed in a way that produces flat images without background subtraction, as shown for the ERO data \citep{EROData}. We compared the noise properties of \Euclid images processed using the LSB method \citep{EROData}, to the official \Euclid pipeline and found them to be similar. Therefore, we subtracted the background determined by the official pipeline from each SC8 exposure before \referee{injecting our clusters and} combining the images into the mosaic. These mosaics contain the uncertainties associated with correction of all of the background components simulated in the mocks and have similar noise properties to the images that would be processed according to the LSB-optimised algorithm with subsequent background subtraction on a scale that is larger than \referee{(and therefore preserves)} the ICL.}

Next, we used our noise images to create weight maps from the inverse variance, using the data quality frame (from the SC8 simulation) to identify bad pixels and set their weights to zero. We finally used \texttt{SWARP} to create weighted mean  \HE mosaics of the four exposures of the ROS. We formatted the results to replicate the SC8 \Euclid mosaics of the ROS and converted the \texttt{SWARP} output weight map back into a noise image. Examples of a simulated cluster \HE mosaic are shown in Fig.\,\ref{fig:mock_redshifts}.  

\subsection{Varying cluster and ICL parameters}
\label{sec:icl_params}

We explore a multivariate cluster and ICL parameter space in order to investigate the effects of different cluster masses and ICL properties, as well as redshift of observation, on our ability to detect the ICL. Since simultaneously varying each parameter would produce an impractically large dataset, we varied each parameter individually, while holding all other parameters fixed at their fiducial values.
In Table\,\ref{tab:parameters} we show a summary of the parameters that were varied and the range of each parameter; the fiducial values of the simulated clusters are shown in bold font.

\begin{table}[]
\caption{Values of halo mass ($\Mhalo$), redshift ($z$), ICL fraction ($\fICL$), effective radius ($r/r_{\mathrm{e},\mathrm{scaled}}$), and S\'ersic index ($n$) that were varied for each of the mock clusters}
    \label{tab:parameters}
    \centering
    \begin{tabular}{c | c | c | c | c}
    $\Mhalo/\Msun$ & $z$ & $\fICL$ (\%) & $r_\mathrm{e}/r_{\mathrm{e},\mathrm{scaled}}$ & $n$ \\\hline
    $\bm{10^{14.7}}$ & \textbf{0.3} & 1 & 0.5 & 0.38 \\
    $10^{14.4}$ & 0.6 & 5 & 0.75 & 0.5\\
    $10^{14.2}$ & 0.9 & 10  & \textbf{1.0} & \textbf{0.76}\\
    $10^{14.1}$ & 1.2 & \textbf{15} & 1.25 & 1.14\\
    $10^{13.9}$ & 1.5 & 20 & 1.5 & 1.52\\
    & 1.8 & 25 &
    
    \end{tabular}
    \tablefoot{Numbers shown in bold indicate the fiducial values that were fixed while other parameters were varied. $r_{\mathrm{e},\mathrm{scaled}}$ is the scale radius of the ICL at a given cluster's stellar mass, defined in Eq.\,\ref{eq:scaling_re}.}
    
\end{table}

In order to explore the effect of cluster halo mass on the ICL detectability, we selected five clusters from the MAMBO catalogue with masses from $10^{13.9}$ to $10^{14.7}\,\Msun $. The upper limit in the mass range corresponds to the most massive cluster in MAMBO; there are no more massive clusters due to the limited volume of the lightcone. We selected the highest mass as the fiducial value since the large size of the EWS means such massive clusters will be plentiful in the survey.

The ICL fraction, $\fICL$, was varied for each simulated cluster across the range of values: 1\%, 5\%, 10\%, 15\%, 20\%, and 25\% by varying the ICL flux input into \texttt{galsim}. The ICL fraction, $\fICL$, is defined as $\fICL = F_{\rm ICL} / (F_{\rm ICL} + F_{\rm BCG} + F_{\rm sat})$ where $F_{\rm ICL}$, $F_{\rm BCG}$ and $F_{\rm sat}$ are the total fluxes of the ICL, BCG and cluster satellite galaxies respectively.

Measurements in the literature display significant scatter in the ICL fraction at $z<1$, \citep{joo2023, contini2024}. \citet{ragusa2023} find an ICL fraction around 15--25\% in clusters with $z \leq 0.05$. At higher redshifts, $1<z<2$,  \citet{joo2023} measure the mean ICL fraction to be ${\sim}\,17\%$, finding no appreciable evolution with redshift.
The ICL fraction has also been found to be linked with halo concentration in semi-analytic models \citep{contini2024b}. Since the concentration-mass relation evolves with redshift, this will have some impact on the ICL fraction, however the scatter in these relations is high and the effect is smaller than the range of our modelled ICL fractions.

Our selected range of ICL fractions is therefore consistent with the typical range of measured ICL fractions in both relaxed and unrelaxed clusters in the literature \citep{krick2007,montes2018,jimenez-teja2018}, with the fiducial value of 15\% chosen to be close to the average ICL fraction measured in clusters between $0<z<2$ \citep{joo2023}.

The shape of the ICL profile was also varied independently, altering the effective radius by up to 50\% of its original value and the S\'ersic index from 0.5--2 times its original value to reflect the ranges of these parameters measured in \citet{kluge20}.

The simulated clusters were drawn from a narrow range of redshifts in the MAMBO catalogue. In order to investigate the effect of redshift, we reposition each cluster across a range of redshifts: $z=\{0.3,\,0.6,\,0.9,\,1.2,\,1.5,\,1.8\}$. The fiducial value was set at $z=0.3$. The angular sizes of each satellite galaxy, the BCG and the ICL were all modified, as was the positional offset of each galaxy from the BCG, according to the corresponding angular size scale at the new redshift. The observed fluxes of all cluster galaxies (and hence the BCG and ICL via the specified $\fICL$) were varied according to the distance modulus and K-correction of the selected redshift.
We do not include any evolution corrections to the clusters or their member galaxies. Instead we are simulating the same clusters observed at different cosmological distances. This means that our mock clusters at higher redshifts do not correspond to the progenitors of the clusters simulated at lower redshifts. We chose this approach so that we can evaluate variations in the S/N arising solely from redshift effects, without the complication of evolution. This also allows us to study the detectability of the most massive clusters at high redshift, even though they are missing from the MAMBO volume. In our analysis, we account for the probability of actually observing such massive clusters by considering the redshift-dependent cluster mass function.

Omitting the effects of evolution does introduce some biases to the observations. Our simualted ICL at high redshift is older, and therefore redder, than it would be in observations. Since older stars have a higher mass-to-light ratio, the true ICL at $z>0.6$ is expected to be slightly brighter than our predictions. The fraction of ICL does not have a strong dependence on redshift \citep{werner2023, joo2023}, so we do not expect this to bias our results. Since the sample of clusters was drawn from the MAMBO catalogue at low redshifts, we expect negligible bias at the low end of our redshift range.

To create the sample of clusters at the desired redshifts, we shift the apparent magnitudes \eceb{listed in the MAMBO catalogue} from the original redshift ($z_{\rm o}$) to the new redshift ($z$). The correction for the distance modulus is straightforward, however a K-correction is also required to account for the fact that each filter will be observing a different region of the rest frame galaxy spectral energy distribution (SED). 
We used \texttt{kcorrect} \citep[v5.0.1,][]{blanton2007} to estimate this correction, \eceb{using custom response curves for each of the \Euclid filters. Since \YE, \JE, and \HE\ will cover bluer wavelengths at high redshifts, we provide \texttt{kcorrect} with the mock photometry from the MAMBO catalogue} for the bands closest to $\lambda_{\HE}\,(1+z_{\mathrm o})/(1+z)$. \texttt{kcorrect} fits the coefficients describing each galaxy's SED, which are then used to apply two K-corrections: once to calculate the rest frame absolute magnitudes, and again to obtain the new observed frame apparent magnitude in the \HE\ band. Examples of a mock cluster transformed to several different redshifts are shown in Fig.\,\ref{fig:mock_redshifts}.

\section{ICL signal-to-noise ratio profiles}
\label{sec:SNProfiles}

\begin{figure}[tbp]
    \centering
    \includegraphics[angle=0,width=\hsize]{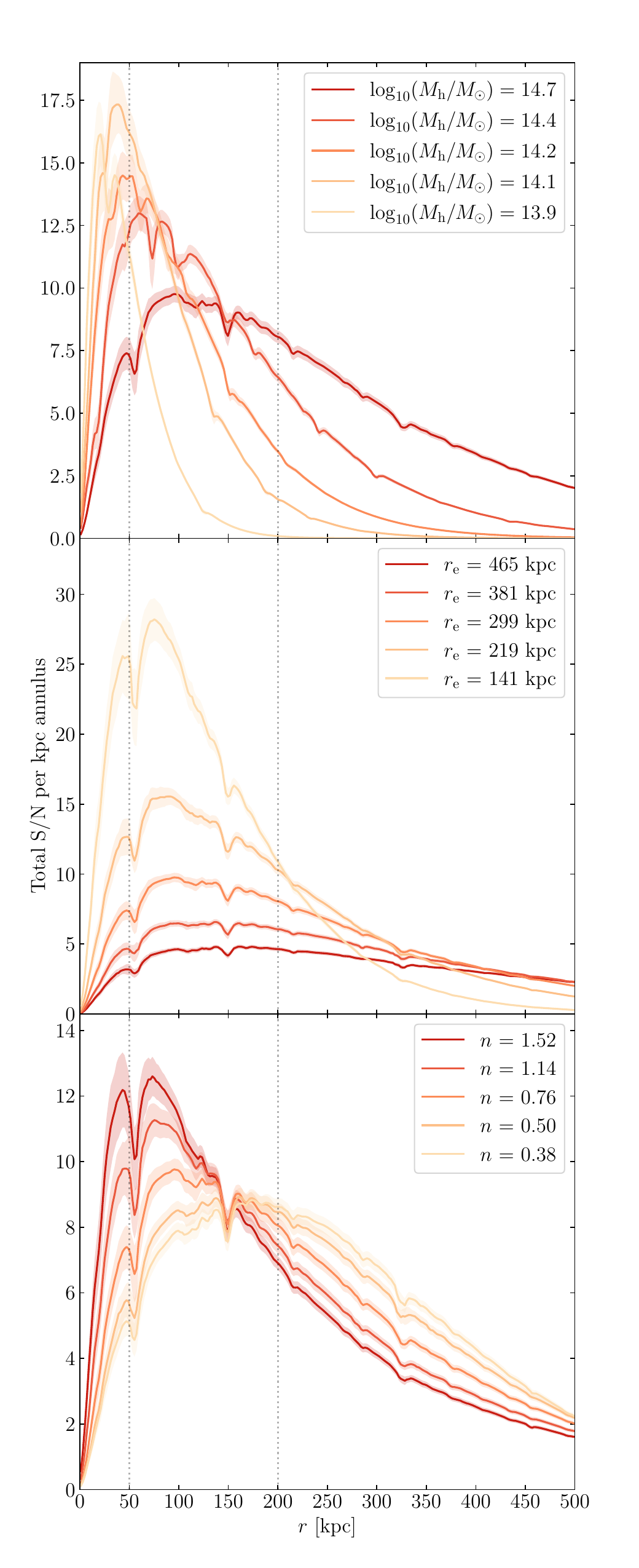}
    \caption{S/N per kpc annulus versus radius for clusters of varying halo mass (\textit{top}), ICL effective radius (\textit{middle}) and ICL S\'ersic index (\textit{bottom}), measured on the isolated ICL component (not including BCG) within 2\,kpc annuli. The lines and shaded regions respectively show the median value and standard deviation of the S/N across nine instances of the same cluster simulated in different regions of the image. \referee{The vertical dotted lines indicate the 50--200~kpc region we selected for the annulus}.
    }
    \label{fig:sn-curve-multi}
\end{figure}

\begin{figure}
    \centering
    \includegraphics[width=\hsize]{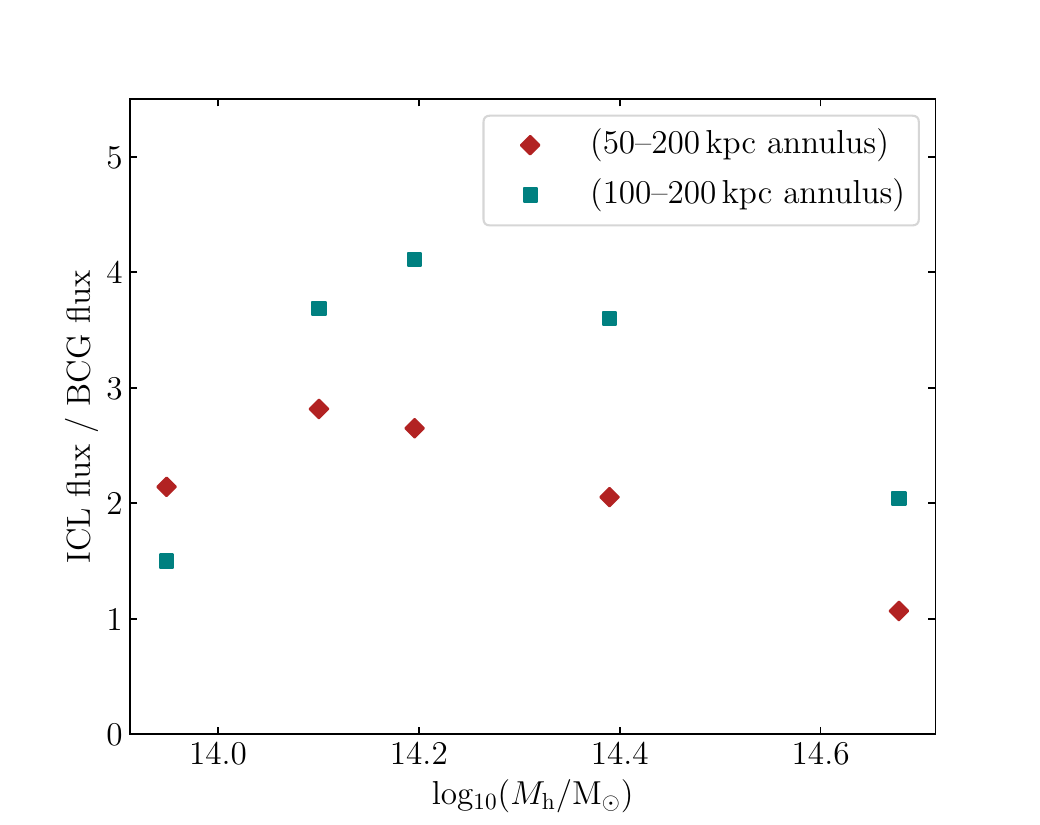}
    \caption{Ratio of the fluxes contributed by the ICL and BCG components of the model clusters, measured within both the 50--200\,kpc annulus (red points) and a 100--200\,kpc annulus (blue points) as a function of halo mass. Each point shows, for a given cluster, the total ICL flux within the annulus divided by the total BCG flux within the annulus.}
    \label{fig:icl_bcg_ratio}
\end{figure}
To determine \Euclid's ability to detect the ICL, we require a measure of the significance of the ICL detection. We opted to use the total S/N within fixed circular annuli, defining the `signal' as the flux of our ICL model summed over all the pixels in the annular aperture. The `noise' is the total uncertainty on that summed flux, which includes a number of contributions.

Firstly, there is the Poisson noise of the model ICL, BCG, and cluster galaxies in each pixel. These are added in quadrature with the \eceb{ systematic noise from the SC8 simulated images into which we inserted our clusters, which itself includes noise from all other sources including zodiacal light, astronomical and instrumental background inhomogeneities, as well as instrumental contributions. To calculate the systematic noise, we measured the standard deviation of the total flux in 500 randomly placed annular apertures over the SC8 simulated image (after masking high surface brightness sources with MTObjects/\referee{Sourcerer} \citealt{mtobjects} using the default parameters). The effects of imperfectly masking fore- and background sources are therefore taken into account in this calculation. We do not model the effects of satellite masking in this analysis, but instead include the noise contributed by satellite galaxies in the Poission noise component.}
In reality, most ICL detection methods \eceb{mask or model and subtract satellite galaxies} before modelling the low surface brightness diffuse light, and will therefore excise these sources of noise. However, masking adds some uncertainty due to both the reduction in effective aperture size and the potential for leaving behind the faint outer regions of masked sources; there is always a compromise to be struck between these competing issues. We therefore expect our uncertainties to be similar to those if we had implemented masking \eceb{of the satellites}.

We remind the reader that our mock images include a variety of observational effects (see Sect.\,\ref{sec:clustersims}), but neglect NISP detector charge persistence and Galactic cirri. The estimated background uncertainty therefore assumes that persistence has been accurately modelled and subtracted, and that the cluster is in a region clear of Galactic cirri.
 
For each simulated cluster, we measure the total S/N of the ICL (i.e., not including flux from the BCG) within circular annuli of width 2\,kpc, extending from the location of the BCG out to 500\,kpc. The S/N of the ICL as a function of radius for varying halo mass, ICL effective radius and ICL S\'ersic index are shown in Fig.\,\ref{fig:sn-curve-multi}. Plots for redshift and ICL fraction are omitted; varying those parameters changes the height of the peak but not the radial location.

\begin{figure*}[tbp]
\centering
    \includegraphics[angle=0,width=\hsize]{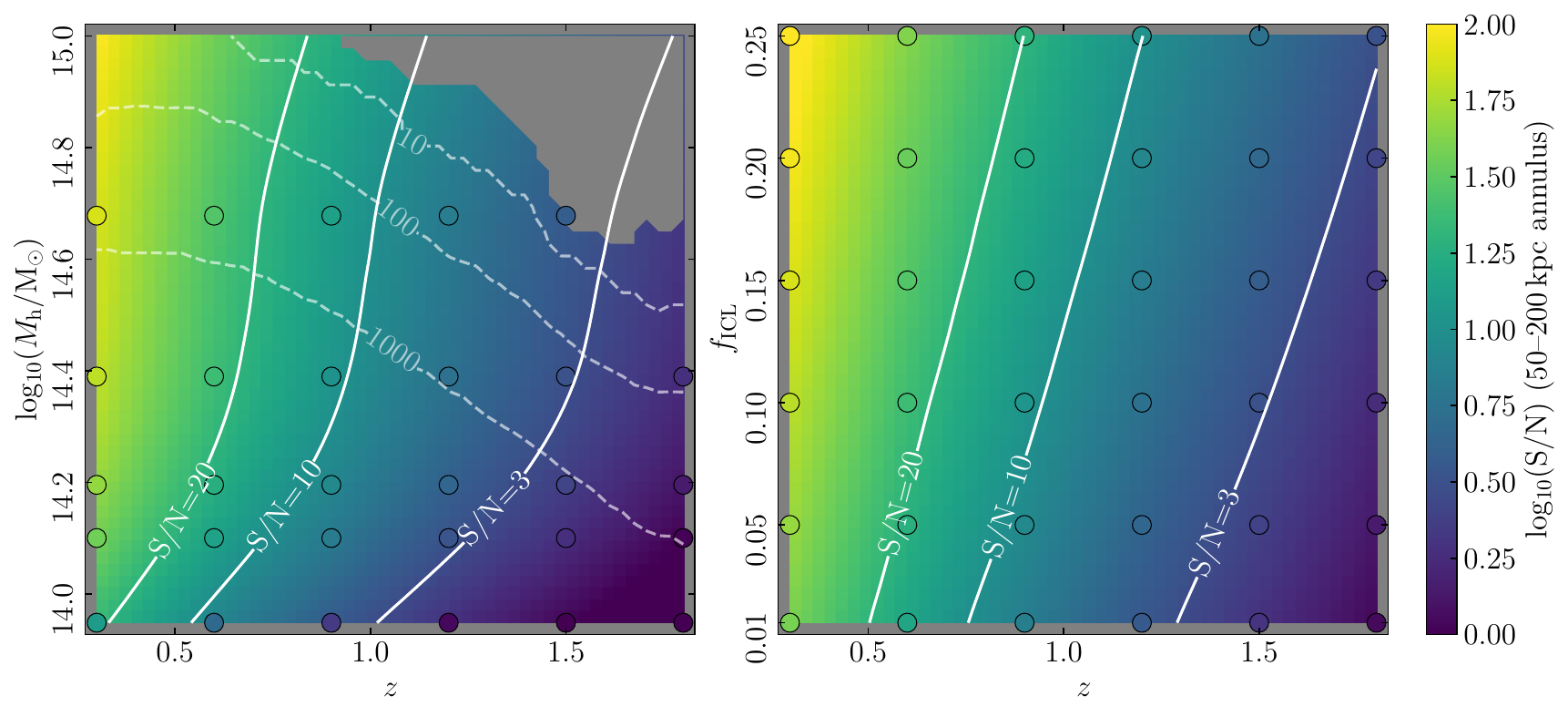}
\caption{\textit{Left}: \HE-band S/N of the ICL + BCG measured within a 50--200\,kpc annulus, interpolated across halo mass and redshift space, using `smoothed' ICL parameters calculated as described in Sect.\,\ref{sec:fitting_sn} assuming a fixed ICL fraction of $15\%$. Coloured circles indicate the S/N values from the individual model clusters for comparison. Contours indicate the lines corresponding to a $\SN= 3, 10$ and 20 as labelled. Dashed contours indicate the threshold halo mass above which the total number of clusters in a redshift bin of $\Delta z=0.1$ is expected to be 10, 100, and 1000, across the footprint of the EWS, calculated as described in Sect.\,\ref{sec:haloestimates}. By the same estimates, the grey shaded region to the upper-right indicates the parameter space in which we expect to find no clusters according to the halo mass function.
\textit{Right}: \HE-band S/N of the BCG+ICL measured within a 50--200\,kpc annulus, interpolated across ICL fraction and redshift space, for a cluster with $\Mhalo=10^{14.7}\,\Msun$. $\fICL$ is defined as the total flux of the ICL across its entire radial extent, divided by the combined total fluxes of the ICL, BCG and cluster galaxies. Circles indicate the locations at which the ICL S/N has been measured to produce the interpolated map.}
\label{fig:mhalo_z}
\end{figure*}

Figure~\ref{fig:sn-curve-multi} shows that there is a peak in the ICL S/N which is generally located in the region 50--200\,kpc from the cluster centre. This is because at large radii, beyond 200--300\,kpc, the ICL flux is diminished and the contribution from satellite galaxies and noise from the nuisance sources substantially impacts the S/N, particularly at higher redshifts, lower halo masses, and ICL fractions. At small radii, within ${\sim}\,50$\,kpc, the noise is completely dominated by the Poisson noise from the bright regions of the BCG. Furthermore, although the surface brightness of the ICL is bright in the inner region, there is comparatively little total ICL flux at small radii due to the small surface area of the innermost annuli, so the $<50$\,kpc region can be excluded without significant loss of total ICL S/N. 

In the top panel, the lower mass clusters peak closer to the innermost limit of the 50--200\,kpc region. Whilst some parameters affect the location of the peak more strongly than others, the majority of the ICL signal is consistently found within this region.

This peak in the ICL S/N enables us to define an annulus of 50--200\,kpc \referee{(marked by the dotted vertical lines in Fig.\,\ref{fig:sn-curve-multi})} within which the total ICL S/N is maximised for a wide variety of clusters. In principle, we could opt for an aperture that scales with $R_{200}$. However, we chose to adopt a fixed aperture to avoid variations when comparing between methods and across cluster properties. Otherwise, each measurement would correspond to a different aperture, confounding interpretation. 
Furthermore, while we know the `true' properties for our mock clusters, in real observations these parameters would first need to be estimated which would add another source of uncertainty. For similar reasons we opt for a circular, rather than an elliptical, annulus, noting that \citet{brough24} found little difference in ICL fractions measured when comparing a circular and elliptical profile.

In the following sections we calculate the S/N of the total combined BCG+ICL flux within this optimal annulus of 50--200\,kpc since, as discussed in Sect.\,\ref{sc:Intro}, it is difficult to separate the BCG and ICL through photometry alone. We therefore quantify how much of the total light in this annulus is contributed by the BCG rather than the ICL. Our simple BCG model has a fixed Sérsic index and half-light radius, but varies in total luminosity in proportion to the cluster's total stellar mass. Hence, we create a brighter BCG in the more massive clusters.

Figure~\ref{fig:icl_bcg_ratio} displays the fraction of light in the 50--200\,kpc annulus that is due to ICL only (not including the BCG), relative to the amount of light contributed by the BCG only. We find that the ICL/BCG light ratio within this annulus is dominated by the ICL component for clusters with $\logMhalo<10^{14.7}$. Although the outer BCG light contributes a significant proportion of the flux in this annulus for the most massive clusters, we note that we are always probing the low surface brightness halo of the BCG since the 50\,kpc radius is almost twice the half-light radius of the modelled BCG.  We therefore define the circular annulus of 50--200\,kpc around the BCG as being the optimal region in which to detect ICL in a cluster for the remainder of this paper.
The blue points in Fig.\,\ref{fig:icl_bcg_ratio} indicate that increasing the inner radius of the annulus to 100\,kpc yields a higher ICL/BCG fraction in the more massive clusters, but it is clear from the top panel of Fig.\,\ref{fig:sn-curve-multi} that this annulus is outside the majority of the ICL flux in the least massive cluster in our sample, and would severely diminish the S/N of the ICL measurements in clusters with $\logMhalo<14$. We further assess the impact of selecting a smaller annulus of 100--200\,kpc within which to detect ICL in Appendix~\ref{sec:contamination}.

\section{Forecasts of the S/N of ICL detections with \Euclid.}\label{sec:sngrids}

To forecast the detectability of the ICL across different cluster masses and redshifts, we measured the total S/N of the flux within a fixed annulus of 50--200\,kpc (as motivated in Sect.\,\ref{sec:SNProfiles}) across our sample of simulated clusters. In this section, and for the rest of the paper, we do not separate the BCG flux from the ICL flux as this cannot be reliably done within observations. Therefore, our measurement of the flux within the 50--200\,kpc annulus includes the flux from the BCG and the ICL, and we refer to this flux as \ICLap{} hereafter. \eceb{The noise in this annulus was measured as the Poisson noise of the model ICL, BCG, and satellite cluster galaxies in each pixel, added in quadrature with the systematic noise from the SC8 simulated images. The systematic noise was calculated as the standard deviation of the surface brightness in 500 randomly placed 50--200\,kpc apertures.}

\label{sec:sngrids_mass}

The coloured circles in the left panel of Fig.\,\ref{fig:mhalo_z} show the S/N ratio with which we can detect \ICLap{} for 30 simulated clusters across a range of halo masses and redshifts, but with $\fICL$ and $n$ fixed at the fiducial values of 15\% and 0.76, respectively. To extend our analysis to regions of higher halo mass not covered by the MAMBO simulations we extrapolate our individual cluster measurements to higher halo masses \referee{using linear fits to cluster properties as a function of mass}.
The \referee{full} details of how we extrapolate to higher masses are described in Appendix~\ref{sec:fitting_sn}.

In addition to extrapolating the results to higher masses, our method of fitting the cluster parameters, described in Appendix~\ref{sec:fitting_sn}, allows us to homogenise the cluster sample by smoothing over variations in individual cluster parameters. The clusters we selected from the MAMBO catalogue exhibit some variation in their luminosities and stellar masses, due to the well-studied intrinsic scatter in the cluster stellar mass-halo mass relation in observations and simulations \citep[e.g.,][]{zu16, golden-marx2018, kravtsov18, pillepich2018}. Combined with the varying geometry of the bright satellite galaxies in each cluster, this results in slight scatter in the measured \ICLap{} S/N values between individual clusters. Correcting for the scatter in the individual cluster measurements therefore allows us to homogenise our ICL properties regardless of individual clusters' variation in stellar mass or luminosity.

Finally, \referee{for visualising the S/N values which lie between those measured in the discrete sample of homogenised, extrapolated clusters,} we apply a 2-D cubic interpolation to the \ICLap{} S/N measurements from the \referee{model} clusters in order to produce the continuous map of S/N across the $\logMhalo$, $z$ space shown in the left panel of Fig.\,\ref{fig:mhalo_z}.
This figure illustrates that with an optimal detection method one could detect (with $\SN\ge 3$) the ICL in most clusters with $\logMhalo \ge 14.4$ at $z\leq1.5$. Similarly, the \ICLap{} flux can be measured with an accuracy of ${\sim}\,10\%$ ($\SN\sim10$) for $\logMhalo \ge 14.4$ and $z\leq 1.0$. 

\begin{figure*}
    \centering
    \includegraphics[width=0.99\textwidth]{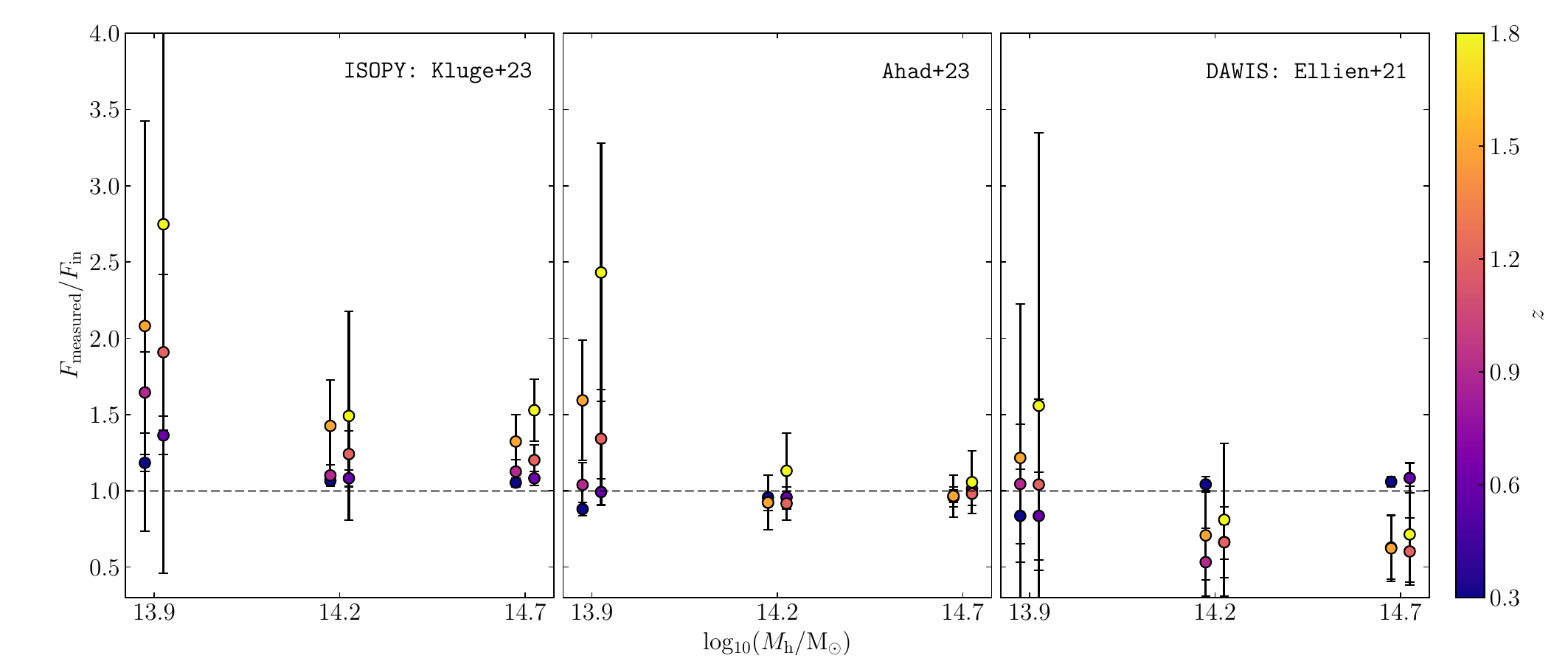}
    \caption{Ratio of the \ICLap{} flux \referee{($F_\mathrm{measured}$)} measured by each fitting method to the \ICLap{} flux of the input model (\referee{$F_\mathrm{in}$, equivalent to the signal component in our S/N measurements)}, for varying $\Mhalo$ and a fixed ICL fraction of $15\%$. Points are coloured by redshift. The errorbars show the standard deviation of all successful measurements of the \ICLap{} flux, divided by the \ICLap{} flux of the input model. Alternating points at each mass and $\fICL$ value have been offset to the right, for clarity.}
    \label{fig:iclfits_comparison_flux_mass}
\end{figure*}

\begin{figure*}
    \centering
    \includegraphics[width=0.99\textwidth]{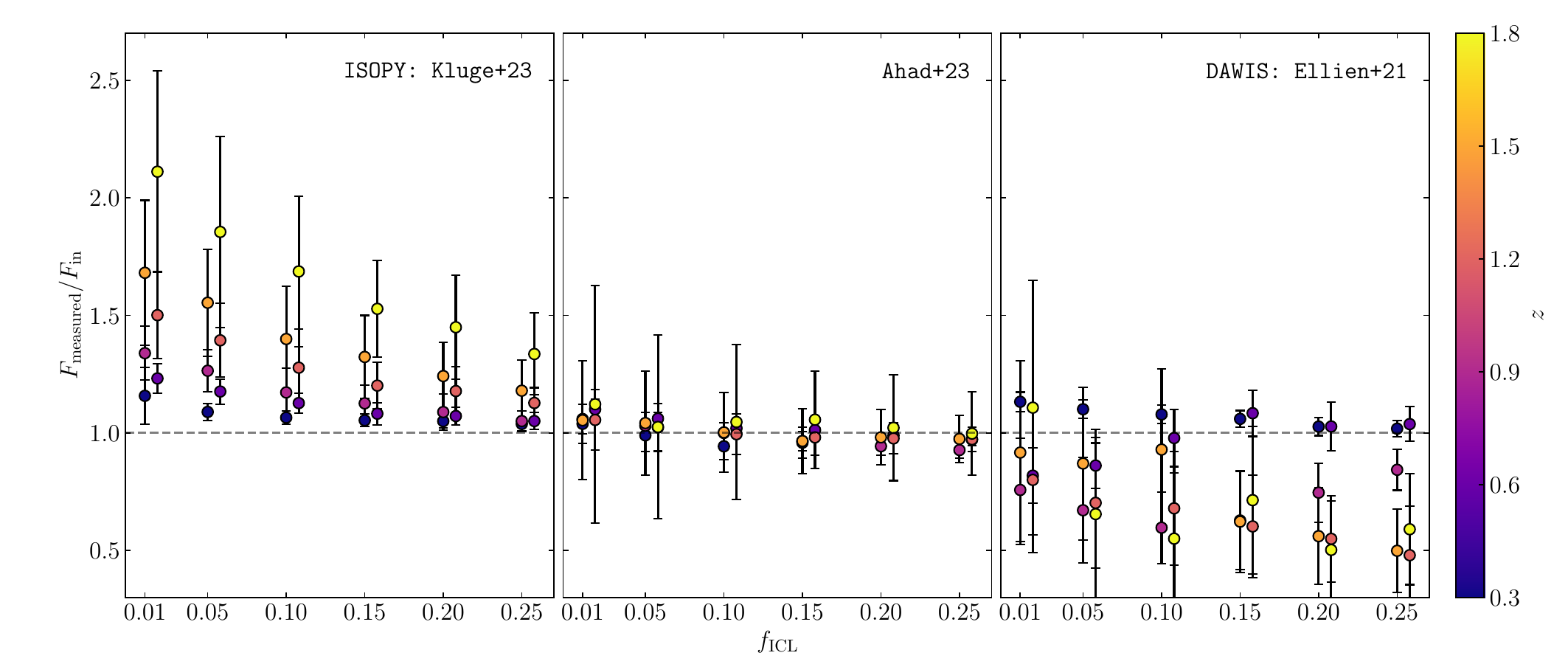}
    \caption{Ratio of the \ICLap{} flux \referee{($F_\mathrm{measured}$)} measured by each fitting method to the \ICLap{} flux of the input model (\referee{$F_\mathrm{in}$, equivalent to the signal component in our S/N measurements)}, for varying $\fICL$ at a fixed halo mass of $\Mhalo=10^{14.7}\,\Msun$. Points are coloured by redshift. The errorbars show the standard deviation of all successful measurements of the \ICLap{} flux, divided by the \ICLap{} flux of the input model. Alternating points at each mass and $\fICL$ value have been offset to the right, for clarity.}
    \label{fig:iclfits_comparison_flux_ficl}
\end{figure*}

\begin{figure*}
    \centering
    \includegraphics[width=0.99\textwidth]{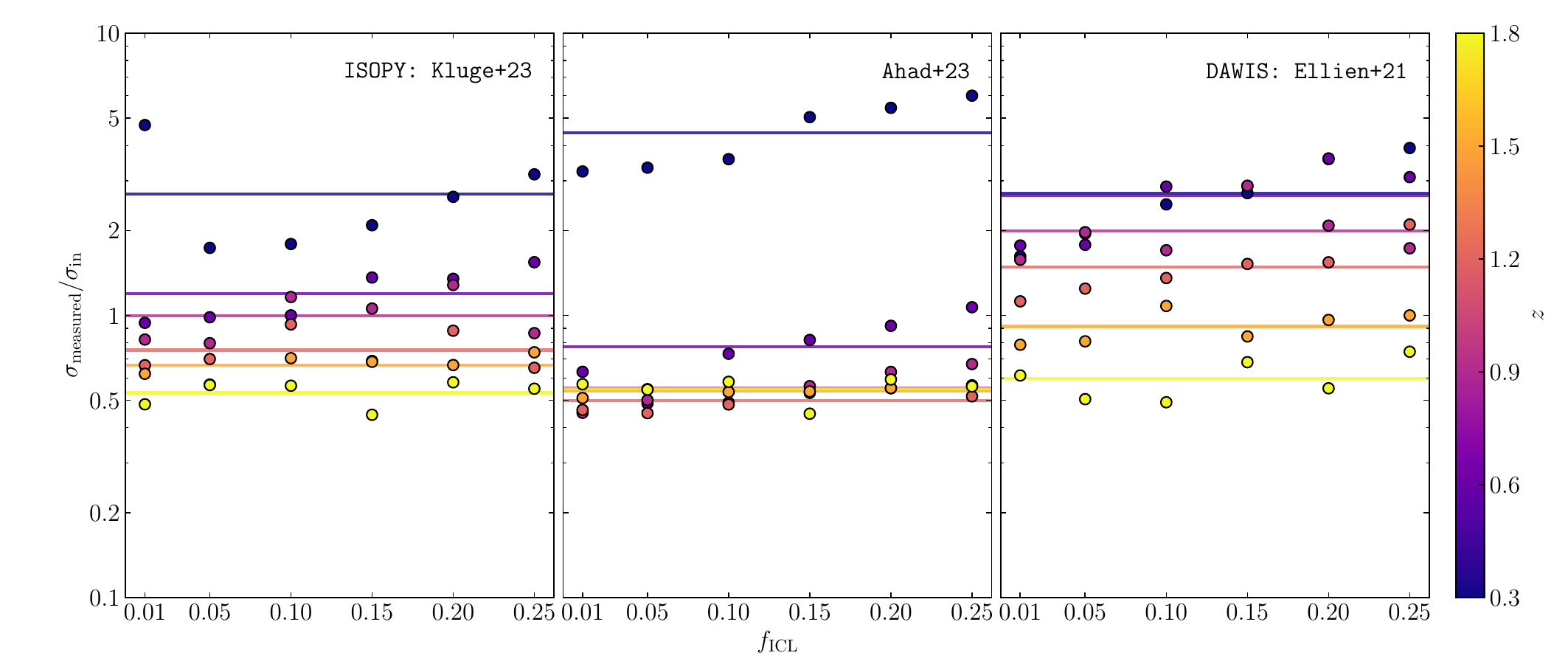}
    \caption{Ratio of the \ICLap{} uncertainty \referee{($\sigma_\mathrm{measured}$)} of each fitting method to the \ICLap{} noise (\referee{$\sigma_\mathrm{in}$, equivalent to the noise component in our S/N measurements)} of the input model, for varying $\fICL$ at a fixed halo mass of $\Mhalo=10^{14.7}\,\Msun$. Points are coloured by redshift. The horizontal lines show the mean value at each redshift, coloured by the corresponding redshift.}
    \label{fig:iclfits_comparison_sigma_ficl}
\end{figure*}

The above results assume a fixed ICL fraction of $15\%$, however the ICL fraction is known to be distributed over a wide range of values \citep{montes2018,jimenez-teja2018}. Measuring this distribution in ICL fraction over a range of redshifts and halo masses is a goal of ICL studies with \Euclid. We therefore explore the effect of the ICL fraction on the measured \ICLap{} S/N.

For this experiment, we fix the halo mass of the cluster to be $\logMhalo=10^{14.7}$, and $n$ is fixed to the fiducial value of 0.76. We simulate 36 clusters at different redshifts and with different ICL fractions. The circles in the right panel of Fig.\,\ref{fig:mhalo_z} display the S/N of \ICLap{} as a function of ICL fraction and redshift for each of the 36 simulated clusters. Since a single cluster was used for this analysis and only $\fICL$ was varied, the measurements are not subject to differences in individual cluster properties as was the case with the left panel of Fig.\,\ref{fig:mhalo_z} and we therefore do not need to use the extrapolated, homogenised cluster measurements. We apply a 2-D cubic interpolation between the \ICLap{} S/N measurements of each cluster to produce the continuous map of \ICLap{} S/N across the $\fICL$, $z$ parameter space.

At redshifts $z\leq0.75$, we expect to be able to observe the \ICLap{} to a theoretical S/N of 10, even if the ICL fractions are only $1\%$ of the total cluster luminosity. However, above $z=1$, the minimum ICL fraction we can observe increases sharply: ICL present at a 1\% $\fICL$ can only be detected with $\SN>3$ at $z<1.3$, for clusters with a mass greater than $\logMhalo=14.7$.

The S/N of \ICLap{} appears to vary more strongly with redshift than with the $\fICL$. This is the result of two effects. Firstly, the BCG contributes some of the light within the 50--200\,kpc aperture, and this BCG contribution does not depend on $\fICL$. This means the flux within the 50--200\,kpc is only partially dependent on $\fICL$. The effect of the BCG contribution on the S/N can be visualised by comparing the right panel of Fig.\,\ref{fig:mhalo_z} with the right panel of Fig.\,\ref{fig:mhalo_z_100_200}. The S/N in Fig.\,\ref{fig:mhalo_z_100_200} is calculated from a 100--200\,kpc annulus which reduces the contribution of the BCG to the derived S/N; as a result, we see that the S/N in the right panel of Fig.\,\ref{fig:mhalo_z_100_200} exhibits more variation with the $\fICL$. Secondly, the variation across the redshift axis has a much larger effect on the flux of the ICL. The redshift evolution varies the ICL flux by the square of the distance, whereas the ICL fraction varies the flux by a factor of 25 at the most.

The S/N values shown here correspond to ideal detection limits, assuming optimised isolation of the BCG+ICL from the background.
In practice the S/N measurements will be somewhat lower due to measurement uncertainties which is explored below.

\section{Comparison of idealised S/N with those from state-of-the-art ICL measurement techniques}
\label{sec:fitcomparison}

The S/N values calculated in the previous section indicate the maximum theoretical S/N attainable assuming optimal masking of satellite galaxies and removal of the background, hereafter referred to as $\SN_{\rm ideal}$. We now consider how these expectations compare with state-of-the-art ICL detection methods, by applying these ICL measurement techniques to our mock cluster images. 

We created several mock images by injecting the same cluster at nine well-separated positions within each \HE\ image. Analysis of all nine realisations of the same cluster allows us to quantify the uncertainties arising from variations in the background, as well as different realisations of the Poisson noise.

We apply three state-of-the-art methods to measure the BCG+ICL flux within each of our simulated images: the masking and isophotal fitting method \referee{ISOPY} \citep{Kluge2023}, a method utilizing \textsc{SExtractor} \citep{bertin1996} to mask satellites and measure the BCG+ICL light in circular apertures \citep{ahad2023}, and a wavelet method: DAWIS \citep{Ellien2021} which decomposes an image into multiple 2D light components at various spatial scales. Each of these methods has different advantages with respect to characterising the details of the ICL, e.g., ellipticity, colour profile, and asymmetry. For the purposes of this paper we focus only on the detection of the ICL. \referee{The different methods also make different choices regarding masking of high surface brightness sources, which could introduce systematic biases.} In Appendix~\ref{isopy_description} we present brief overviews of the different BCG+ICL measurement methods. Each method isolates the BCG+ICL flux, and \ICLap{} is then measured from the resulting 1D \citep{ahad2023} or 2D (\texttt{isopy} and DAWIS) BCG+ICL profiles. For each batch of nine identical clusters we calculate the mean of the detected flux in \ICLap, $F_{\rm measured}$, and the standard deviation of the nine clusters, $\sigma_{\rm measured}$. These are then compared to the input flux in \ICLap\, $F_{\rm in}$ (\referee{equivalent to the signal component in our S/N measurements}), in Figs.\,\ref{fig:iclfits_comparison_flux_mass} and \ref{fig:iclfits_comparison_flux_ficl}, and the ideal noise estimated in Sect.\,\ref{sec:sngrids}, $\sigma_{\rm in}$ (\referee{equivalent to the noise component in our S/N measurements}), in Fig.\,\ref{fig:iclfits_comparison_sigma_ficl}.

In early tests of the ICL measurements, the methods were tuned using the `true' ICL fluxes in similar data in order to optimise their masking algorithms for the \Euclid images. However, for the final measurements presented here, the analysis was carried out `blindly' with no knowledge of the input BCG+ICL fluxes.

The left panels of Figs.\,\ref{fig:iclfits_comparison_flux_mass} and \ref{fig:iclfits_comparison_flux_ficl} show that the \texttt{isopy} \citep{Kluge2023} method tends to measure a higher \ICLap{} flux than that of the input model. This overestimation can be due to a number of reasons, including incomplete masking of high surface brightness sources and underestimation of the background.
The bias is greater for clusters at higher redshifts compared to lower redshifts: a factor of ${\sim}\,1.5$ at $z=1.8$ compared to a factor of $1.1$ at $z=0.3$ \referee{in the case of the fiducial cluster}. There is a small trend with $\fICL$ in which clusters with lower $\fICL$ have a  higher bias than clusters with higher $\fICL$. For clusters with masses $\logMhalo\ge14.2$ the bias does not depend on mass, but the bias is  higher in the lowest mass clusters in our simulations.

The middle panels of Figs.\,\ref{fig:iclfits_comparison_flux_mass} and \ref{fig:iclfits_comparison_flux_ficl} show that the \ICLap{} flux measured by the \citet{ahad2023} method exhibits little bias at all redshifts for clusters with masses $\logMhalo\ge14.2$. The results also have fairly low scatter between the measurements of the 9 instances of the cluster, and are consistent across the redshift range. Similar to the \texttt{isopy} results, the method overestimates the \ICLap{} flux for clusters of $\logMhalo=13.9$. This is likely driven by Eddington bias, with very large scatter between the measurements especially at higher redshifts.
The middle panel of Fig.\,\ref{fig:iclfits_comparison_flux_ficl} reveals that there is no detectable trend in bias with ICL fraction.

The results from \texttt{DAWIS} \citep{Ellien2021}, in the right panel of Fig.\,\ref{fig:iclfits_comparison_flux_mass}, show that for low redshifts and higher cluster masses, the bias is minimal, and the method retrieves close to the input \ICLap{} flux. At higher redshifts, the method slightly underestimates the \ICLap{} flux. For lower-mass clusters, whilst all of the instances of the cluster yielded a measurement, the uncertainty in the results across each of the 9 measurements is significantly higher, and the measured fluxes at each redshift show significant scatter around the input value, ranging from ${\sim}\,80\%$ of the measured \ICLap{} flux at $z=0.3$, to ${\sim}\,160\%$ at $z=1.8$.
The right panel of Fig.\,\ref{fig:iclfits_comparison_flux_ficl} shows that there is very little dependence of the bias on the ICL fraction, with the measurements at all $\fICL$ values underestimating the \ICLap{} flux at high redshifts.
This underestimation may be due to the fixed wavelet separation used to extract the ICL regardless of mass and redshift, and may be corrected by fine tuning.

This systematic bias is consistent with the findings of \citet{brough24} who investigated a larger variety of ICL detection methods, which included the \citet{ahad2023} and \citet{Ellien2021} methods tested here. The systematic bias of the detection methods can be estimated and removed when the bias is driven by incomplete or overzealous masking of high surface brightness sources and under/overestimation of the  background. This can be done by performing a set of simulations of clusters with ICL, as we have done here, and determining the flux correction factors that must be applied; this task will be performed in a future work when analysing the on-sky data releases of \Euclid. Systematic biases are much harder to fix when the root cause is Eddington bias as this cannot be done for individual clusters, but only on large samples of clusters. This suggests that individual ICL measurements made in the EWS of clusters with $\logMhalo\le 14.2$ at $z>0.6$ should be treated with caution. In terms of the main goal of this paper, which is to forecast the detectability of ICL in clusters with \Euclid, this result suggests that we cannot reliably detect ICL in clusters with halo masses of $\logMhalo\le 14.2$ at $z>0.6$ because of the significant impact of Eddington bias. We note that Fig.\,\ref{fig:sn-curve-multi} shows that our 50--200\,kpc aperture is not optimised for the S/N of clusters with $\logMhalo\le 14.2$, and a smaller, more compact aperture of 30--100\,kpc may avoid the Eddington bias issue for these low mass clusters.

For clusters with $\logMhalo\gtrsim14.2$, the systematic bias will be measured and corrected in the EWS. Therefore, the S/N of the \ICLap{} measurements by each of the methods is $F_{in}/\sigma_{\rm measured}$. Hence, Fig.\,\ref{fig:iclfits_comparison_sigma_ficl} shows the ratio of $\SN_{\rm measured}$ to $\SN_{\rm ideal}$. For instance, a ratio of 2 in $\sigma_{\rm measured}/ \sigma_{\rm in}$ means that the detection method would measure a S/N that is half of the idealised S/N shown in Fig.\,\ref{fig:mhalo_z}.
\eceb{Note that the values in Fig.\,\ref{fig:iclfits_comparison_sigma_ficl} are sometimes less than 1, which implies that the results of the measurement methods have, in some cases, lower uncertainties than the noise in the data. This is because the total simulated noise $\sigma_{\rm in}$ incorporates the systematic noise due to variation of the background flux on 50--200\,kpc scales across the image, which varies due to fluctuations on both small and large ($>500$\,kpc) scales. Each of the measurement methods performs an estimation and removal of the large-scale background around each cluster in the nine pointings, which therefore removes the scatter caused by large-scale fluctiations and thus reduces the scatter between the measured fluxes. Since ICL will be measured from the \Euclid images using similar methods, $\SN_{\rm measured}$ will be more representative of the uncertainty of the ICL measurements.}

The left panel of Fig.\,\ref{fig:iclfits_comparison_sigma_ficl} shows that the $\SN_{\rm \texttt{isopy}}$ is between 3 times lower and 2 times higher than $\SN_{\rm ideal}$, with a strong dependence on the redshift of the cluster. There is very little dependence on $\fICL$ at $z\geq 0.6$, and the mean change in $\SN_{\rm isopy}$ below $\SN_{\rm ideal}$ is a factor of 1.2, 1.0, 0.8, 0.7 and 0.5 at $z=0.6$, 0.9, 1.2, 1.5, and 1.8 respectively.
We also find that there is minimal difference in these factors for clusters of different halo masses $\logMhalo\geq 14.2$, and therefore do not show results for halo masses other than $\logMhalo = 14.7$ in Fig.\,\ref{fig:iclfits_comparison_sigma_ficl}. The situation is different for nearby clusters at $z=0.3$: there is a strong dependency on $\fICL$ with the decrease in $\SN_{\rm isopy}$ below $\SN_{\rm ideal}$ ranging from a factor of 2 at $\fICL=0.05$ to 3 at $\fICL=0.25$. \eceb{Whilst this indicates a decrease in S/N}, the right panel of Fig.\,\ref{fig:mhalo_z} 
shows that $\SN_{\rm ideal}$ for all of these clusters ranges from $50$ at $\fICL=0.05$ to $\sim100$ at $\fICL=0.25$. Hence, even with a reduction in the S/N by factors of 2--3, we will still reliably detect all clusters at these masses at $z\sim0.3$ with a $\SN>16$.

In the middle panel of Fig.\,\ref{fig:iclfits_comparison_sigma_ficl}, the $\sigma_{\rm measured}$ of the 1-D profile method is low for redshifts $z>0.3$, ranging from a factor of 0.5 to 0.8 times the noise of the input model. This means that in the best cases, the ICL measurement introduces no discernable additional uncertainty over the sources of noise we have included in the theoretical model \eceb{and due to the local background estimation, performs better than the total calculated noise across the modelled clusters}. The results at $z=0.3$ exhibit more uncertainty compared with the input model, likely resulting from a combination of the higher flux of the ICL, \referee{an increased number of interloping sources across the cluster's larger angular size}, and the increased challenge of masking satellite galaxies at lower redshifts (the larger angular size of the satellite galaxies at low redshifts means that the measured fluxes are more sensitive to over/undermasking). The $z=0.3$ results also follow a similar trend with $\fICL$ to that observed in the results from \texttt{isopy}.

The right panel of Fig.\,\ref{fig:iclfits_comparison_sigma_ficl} shows that $\SN_{\rm\texttt{DAWIS}}$ ranges from 0.4 to 2 times the $\SN_{\rm ideal}$, with a dependence on $\fICL$ at $z=0.3$ which disappears, or is indistinguishable due to the scatter between points, for $z\geq0.6$. The mean \eceb{factor} in $\SN_{\rm\texttt{DAWIS}}$ ranges from $2$ times lower at $z=0.6$ to $1.6$ times higher at $z=1.8$. Similarly to \texttt{isopy}, we see little variation in these values for different halo masses. These results show that the uncertainties on the measurements yielded using wavelet fitting are higher than those for the other methods. However, a benefit of wavelet fitting is that it is able to detect features and substructures within the ICL, since it does not assume a symmetrical isophotal model.

In summary, \eceb{any of the state-of-the-art ICL measurement algorithms discussed here} are able to measure the \ICLap{} flux to a S/N that is comparable to or even better than $\SN_{\rm ideal}$ for clusters with $\logMhalo\gtrsim 14.2$ at $z\geq0.6$. At lower redshifts, the difference in the S/N is larger, but since $\SN_{\rm ideal}$ is so high for these clusters, \Euclid\ will still be able to detect the ICL in $z\sim0.3$ clusters with $\logMhalo>14.2$ at a typical $\SN>10$.
Wavelet-fitting and isophotal-fitting techniques are able to retrieve the ICL with a \eceb{slightly} lower S/N ratio, but their versatility is better suited to measuring the shape and distribution of the ICL, rather than the initial detection.

In the left panel of Fig.\,\ref{fig:mhalo_z}, the $\SN_{\rm ideal}$ contours do not have a strong dependency on halo mass for $\logMhalo\geq 14.2$ and are almost aligned perpendicular to the redshift axis.\footnote{Regardless of the $\SN_{\rm ideal}$ in Fig.\,\ref{fig:mhalo_z}, state-of-the-art ICL detection methods cannot measure the ICL in clusters at $\logMhalo\geq 14.2$ at $z\ge0.6$ reliably using \ICLap{} since their detection is greatly influenced by Eddington bias, which cannot be corrected for in individual clusters.} Therefore, the S/N contours achieved by state-of-the-art methods are very close to $\SN_{\rm ideal}$.

\section{The expected number of clusters with detectable ICL within the Euclid Wide Survey}
\label{sec:haloestimates}

The above sections describe the \ICLap{} detection limits of clusters with various halo mass, redshift and ICL fraction within the EWS. We now use this information to predict the total number of clusters from which \Euclid will be able to measure ICL by combining our results with an analytical estimate of the cluster mass function \eceb{across the ${\sim}\,14\,000\,\mathrm{deg}^2$ footprint} of the EWS.

We use the semi-analytical approach described in \citet{EP-Sereno} to devise an estimate for the number of clusters with $\logMhalo > 14.2$ in which we expect to measure the ICL above a given $\SN_{\rm ideal}$. We choose to limit our sample to $\logMhalo > 14.2$ because of Eddington bias affecting measurements of \ICLap{} at lower cluster masses.
We note that in this analysis we do not take into account the cluster selection function of \Euclid cluster-finding algorithm, \eceb{which is predicted to have a completeness of $\sim80\%$ at all redshifts $z\leq2$} \citep{sartoris2016}. Hence we derive the maximum possible number of clusters, at various redshifts, that will host detectable levels of ICL.

We take our estimates for the expected $\SN_{\rm ideal}$ at a given halo mass and redshift calculated in Sect.\,\ref{sec:sngrids} and combine these with the halo mass function from \citet{tinker2008}, using the cosmological parameters listed in Sect.~\ref{sc:Intro} ($\Omega_\mathrm{m}=0.3$, $H_0=70\,\mathrm{km}\,\mathrm{s}^{-1}\,\mathrm{Mpc}^{-1}$) and a $\sigma_8$ of 0.8.  We note that all the numbers we quote below will depend on this assumed cosmology. For details on the semi-analytical approach we refer to \citet{EP-Sereno} and \citet{ingoglia24}.

\begin{figure}[tbp]
\centering
\includegraphics[angle=0,width=1.0\hsize]{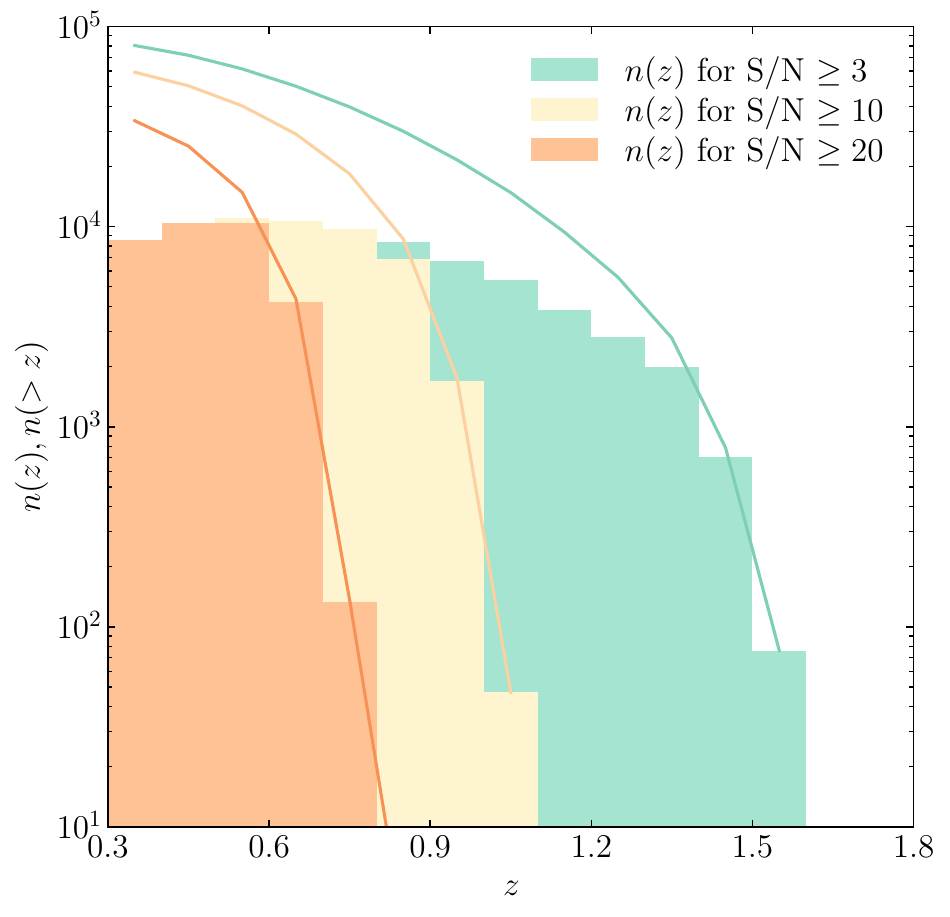}
\caption{The predicted number of clusters above a given \ICLap{} S/N threshold within the footprint of the Euclid Wide Survey, as a function of redshift. Solid bars indicate the number of clusters within a ($\Delta z=0.1$) redshift bin. The lines indicate the cumulative number of clusters above a given redshift. Values are shown for $\SN_{\rm ideal }$ = 3, 10, and 20, calculated using the \citet{tinker2008} halo mass function modelling via the methodology presented in \citet{EP-Sereno}.}
\label{fig:nclust_SN}
\end{figure}

By integrating the halo mass function we obtain predictions for the number of clusters we expect to find above a given $\SN_{\rm ideal}$ of 3, 5, 10 and 20 as a function of redshift, the results of which are shown in Fig.\,\ref{fig:nclust_SN}. The EWS will detect ICL in ${\sim}\,10^4$ clusters per $\Delta z =0.1$ across $0.3<z<0.6$. All of these detections will have a $\SN_{\rm ideal}>20$ (and comparable realistic S/N). For clusters between $z=0.6$ and $z=0.9$, the EWS will detect ICL in $7000$--$10\,000$ clusters per $\Delta z =0.1$, most of which will have a $\SN\ge10$. At the edge of the ICL detection limit, with an ideal $\SN=3$, the EWS will detect ICL in a rapidly declining number of clusters at $z>1.0$: several thousand at $z=1.0$, but less than a hundred at $z\sim1.5$. The rapid drop in the number of clusters with detectable ICL at higher redshifts is driven by the scarcity of massive clusters at these redshifts. ICL will only be detected in less than a hundred clusters at $z>1.5$ with a $\SN\sim3$.  We note that these numbers, and the distribution shown in Fig.\,\ref{fig:nclust_SN}, do not take into account the cluster-selection function of the \Euclid data, and only determines whether the ICL can be detected above a certain S/N. Depending on the parameters of the cluster detection algorithm, it is possible that ICL could be detected in clusters that were not selected by the cluster-finding algorithm. Moreover, features in our Galaxy, such as cirrus and bright stars, may overlap along the line of sight with clusters, effectively making it impossible to isolate and study the ICL. We do not take these features into account in the quoted numbers, therefore the quoted numbers should be considered upper limits of the number of clusters in which we can study ICL.

In total, approximately $80\,000$ clusters in the EWS region will have detectable ICL within a 50--200\,kpc annulus around the BCG. More than 90\% (50\%) of these clusters lie within $0.3<z<1.2$ ($0.3<z<0.75$).
Our results indicate that the optimal redshift range for ICL studies that investigate the relation between ICL properties with cluster properties is $0.3<z<0.75$, whilst the evolution of ICL properties will be measurable out to $z=1.5$. Although our predictions assume an ICL fraction of 15\%, Fig.\,\ref{fig:iclfits_comparison_sigma_ficl} shows that the ICL fraction has little discernible effect on the detectability of the ICL.  

\eceb{One of the primary sources of background contamination mentioned in Sect.~\ref{sec:clustersims} is Galactic cirrus.  Galactic cirrus can be modelled and subtracted from the observations if high-resolution far-infrared data is available \citep{Mihos2017}, but this is not available for much of the EWS. Alternatively, cirrus has recently been successfully modelled based on its morphology and using multi-wavelength colour information, as shown by \citet{Liu2024} and in the ERO observations of A2390 \citep{ellien2025}.}

\begin{figure}[tbp]
\centering
\includegraphics[angle=0,width=1.0\hsize]{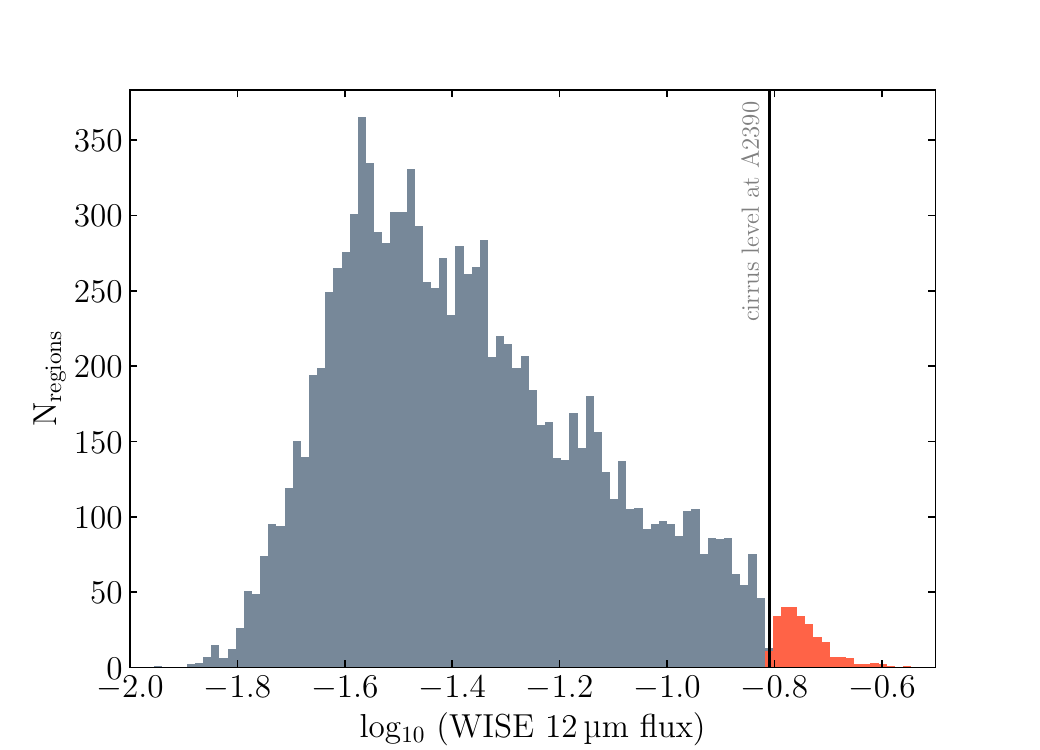}
\caption{Measurements of WISE $\SI{12}{\micro\metre}$ flux \referee{(MJy/sr)} within a 30 arcminute aperture around 5870 randomly selected pointings within the footprint of the DR1 survey. Values greater than that measured within the field of A2390 are shown in red, corresponding to around $2\%$ of the coverage.}
\label{fig:hist_cirrus}
\end{figure}

\eceb{We cannot estimate the fraction of EWS area that is completely free of cirri, since we do not yet have high resolution images of the entire area. Instead, we estimate the effect of cirrus on our total number estimates by comparing the cirrus present in a ``worst-case'' scenario with the cirrus level across the footprint of DR1.
For the upper limit on the acceptable level of cirrus, we take the ERO observations of A2390 (Ellien et al., submitted.) as an example. The level of cirrus present in the field of A2390 is high, but measurements of the ICL are still achievable by careful modelling and subtraction of the cirrus component.
Taking 11581 random pointings across the footprint of the Euclid DR1 coverage, we measured the WISE $12\mu m$ flux within an aperture of 30 arcminutes (\referee{due to the fractal nature of the cirrus, the variation on a 200\,kpc scale, which is the main factor of concern, is expected to be proportional to the large-scale flux of the cirrus}). \referee{The results are shown in Fig.\,\ref{fig:hist_cirrus}}. We found that 248 of the random pointings had a flux equal to or higher than that present in the field of A2390. We therefore estimate that approximately 2\% of the coverage of DR1 will be so greatly impacted by cirrus that it precludes a measurement of the ICL. \referee{The regions strongly affected by cirrus are also the regions where the source density of foreground stars is likely to increase to an extent which may also impact our ability to measure ICL.}}

\section{Characterising the ICL at large radii}
\label{sec:extent}

\begin{figure}[tbp]
\centering
\includegraphics[angle=0,width=1.0\hsize]{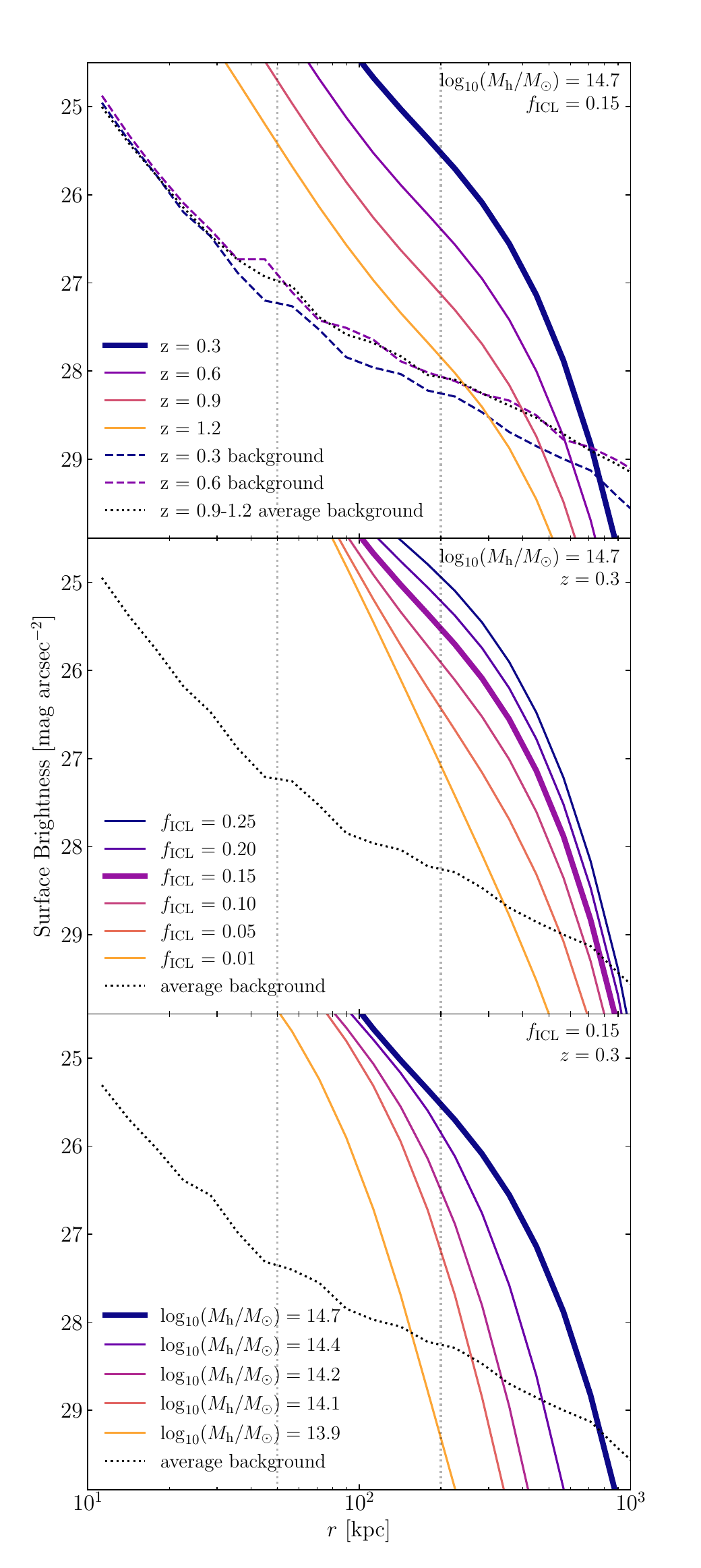}
\caption{The expected surface brightness profile of the BCG+ICL (solid coloured lines) and $3\sigma$ noise  level (dashed lines) for a range of redshifts ($z$; \emph{upper} panel), ICL fractions, ($\fICL$; \emph{central} panel), and halo masses ($\Mhalo$; \emph{lower} panel), varying with respect to our fiducial model: $z=0.3$, $\fICL=0.15$ and $\logMhalo=14.7$. Profiles are shown up to redshift 1.2, above which a cluster of this mass is not expected to exist within the area of the EWS (see Fig~\ref{fig:mhalo_z}, grey shaded region). The fiducial model is indicated by a thick line in each panel. Moreover, the black dotted line in each sub-figure is representative of the background noise level for the fiducial model.  The noise level depends on the redshift of the cluster (see upper panel), but varies little for $z > 1.0$.  The radius to which ICL can be detected increases with $\fICL$ and $\Mhalo$, while decreasing with redshift.}
\label{fig:radial_extent}
\end{figure}

\begin{figure*}[tbp]
    \centering
    \includegraphics[angle=0,width=1.0\hsize]{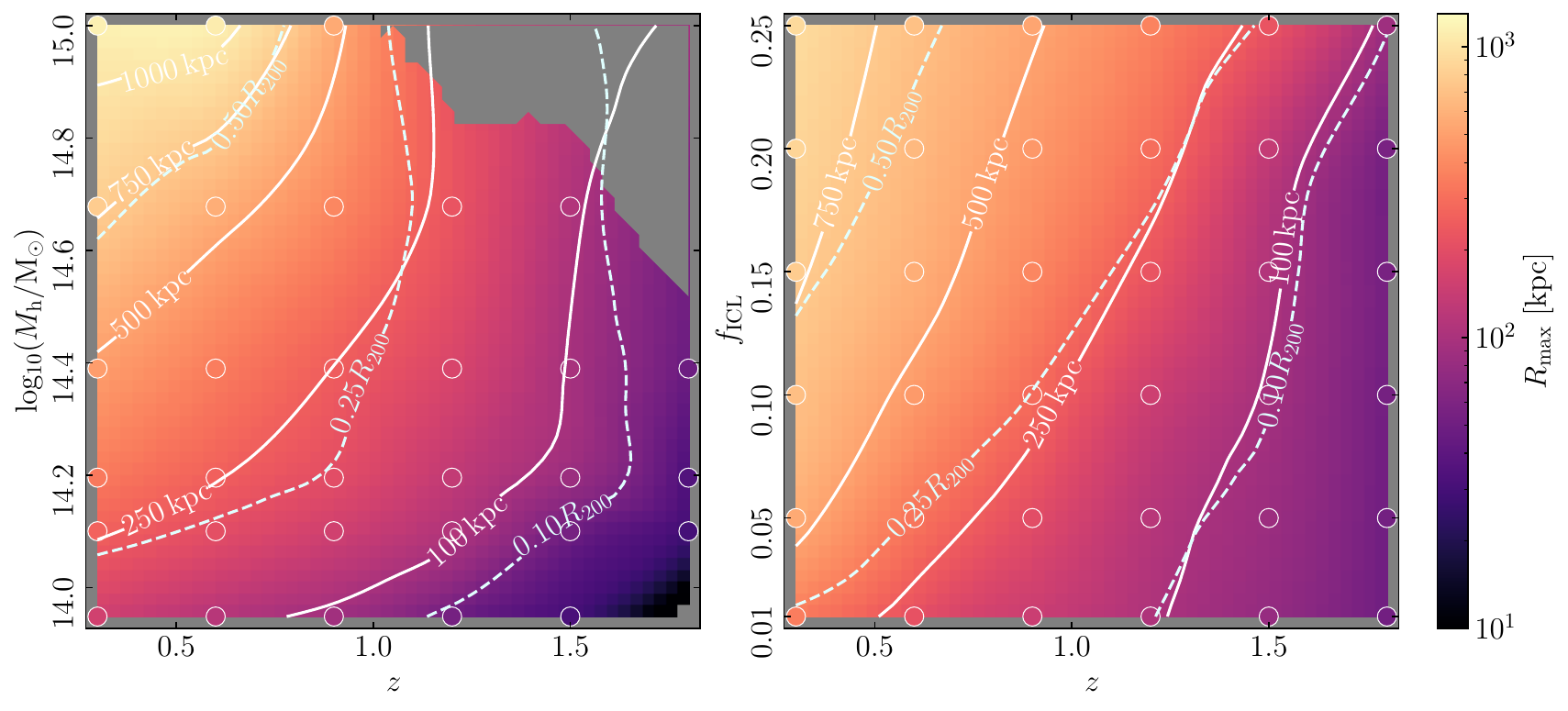}
    \caption{The maximum radius to which the ICL can be detected, $\Rmax$, interpolated across halo mass, $\logMhalo$, and redshift, $z$, for a fixed $\fICL = 0.15$ on the left panel and across ICL fraction, $\fICL$, and redshift, $z$, for a fixed halo mass, $\logMhalo = 14.7$ on the right panel. Coloured circles indicate the measured $\Rmax$ values from individual mock clusters. Contours show the limiting radial extent expressed in physical units by the solid lines and in terms of $R_{200}$ by the dashed lines, where the value of $R_{200}$ corresponds to a typical cluster at the corresponding location in the parameter space.}
    \label{fig:grid_rmax_mass}
\end{figure*}

In this section, we \eceb{explore the S/N of the BCG+ICL as a function of radius from the BCG centre. This will be evaluative of \Euclid's potential for measuring key characteristics of the ICL, e.g. the ICL fraction and BCG offset, and identifying interesting features, such as the splashback radius, all of which require accurate measurements of the ICL flux out to large radii.}

We measure the radial profile of the BCG+ICL surface brightness in a series of circular annuli.
Since the surface brightness of the ICL decreases rapidly with radius, it is common to choose annular bins that increase in width at larger radii. Typically, this width is chosen to be a fixed fraction of the radius, i.e.~a constant in the logarithm of the radius (note that for the purposes of Fig.\,\ref{fig:sn-curve-multi} we used a fixed, linear, annular width). 
We chose a radial spacing of 0.1\,dex for our mock images.

The radial profile of the noise includes contributions from the Poisson noise of the BCG, ICL, and all other astrophysical sources in each annulus, as well as the systematic uncertainty of the background level which dominates the noise profile at large radii. \eceb{The systematic uncertainty of the background within each annulus was measured as the standard deviation of the flux measured in 500 annuli (of the same size) placed in random positions on the SC8 mosaic where all high surface brightness sources were masked.}   We multiply the final radial profile of the noise by three to obtain the $3\sigma$ background level, and show the resulting surface brightness profiles of the noise by the dashed lines in Fig.\,\ref{fig:radial_extent}. These lines should be interpreted as the surface brightness limit to which BCG+ICL can be detected with a $\SN=3$ at a given radius.
Given the steep drop off in ICL surface brightness at large radii (see Fig.\,\ref{fig:radial_extent}) there is only a small  difference in the maximum radial extent to which we can detect ICL for a S/N in the range 3--5.  

The surface brightness profiles of the BCG+ICL clusters are derived from the mean of the nine simulated clusters per set of parameters. In Fig.\,\ref{fig:radial_extent}, we show the impact of varying $\fICL$, redshift, and $\Mhalo$ on the surface brightness profile, but choose to keep the ICL profile shape fixed to $n=0.76$ and limit its size ($r_{\rm e}$) to vary according to Eq.\,(\ref{eq:scaling_re}); see Sect.\,\ref{sec:icl_params} for details. Varying $n$ and $r_{\rm e}$ significantly affects the ICL S/N at large radii, as shown in the lower two panels of Fig.\,\ref{fig:sn-curve-multi}, therefore the surface brightness profiles are only indicative of the average for a population of clusters. Nevertheless, they allow us to roughly estimate the region of parameter space ($z$, $\Mhalo$, $\fICL$) where we will be able to explore ICL properties to appreciable fractions of the cluster radius. From this, we can estimate -- to an order-of-magnitude -- the number of clusters in the EWS with which we will be able to perform such studies.

We determine the maximum radius at which we can detect ICL, $\Rmax$, from where the BCG+ICL surface brightness profile intersects the radially-dependent detection limit. 
In Fig.\,\ref{fig:radial_extent} we have only varied one parameter at a time, but the trends are consistent across the parameter space.  To visualise this more thoroughly, we show $\Rmax$ as a smooth function of $z$, $\Mhalo$ and $\fICL$ in Fig.\,\ref{fig:grid_rmax_mass}. Both panels of the figure express the limiting radial extent in physical units measured for our model clusters by the solid contours. The dashed contours show the same values as a ratio of the typical $R_{200}$ for a typical cluster at the same mass and redshift, calculated using the python package \texttt{cluster-lensing} \citep{ford2016}.

The BCG+ICL of our fiducial cluster ($\logMhalo=14.7$, $\fICL=0.15$, and $z=0.3$) is detected with a $\SN>3$ out to ${\sim}\,770$\,kpc, which is ${\sim}\,0.5 R_{200}$. However, $\Rmax$ declines strongly with redshift. At $z=1.2$, the BCG+ICL of a cluster with the same properties can only be detected to ${\sim}\,220$\,kpc. Note that this cluster is detected with a S/N of 7 in \ICLap.
The corresponding limiting surface brightness of the detection, assuming an ICL fraction of 15\% and a halo mass of $\logMhalo=14.7$ ranges from ${\sim}\,29$\,mag\,arcsec$^{-2}$ at $z=0.3$ to ${\sim}\,28$\,mag\,arcsec$^{-2}$ at $z=1.2$.

The left panel of Fig.~\ref{fig:grid_rmax_mass} also shows that $\Rmax$ depends on the cluster mass such that we can detect the ICL out to significantly greater distances for more massive clusters. Assuming our fiducial values of $\fICL$ and redshift, but for an $\logMhalo=13.9$ cluster, we can only detect the ICL out to 150\,kpc. Although lower mass clusters have a smaller $R_{200}$, this $\Rmax$ corresponds to only 15\% of $R_{200}$. Recall that, for our fiducial halo with $\logMhalo=14.7$, $\Rmax \sim 0.5 R_{200}$. Thus, for more massive haloes, even at fixed $\fICL$, we can not only detect the ICL at larger physical radii, but also out to a greater fraction of the size of the cluster. This behaviour is a consequence of our assumed, but observationally-motivated, dependence of ICL luminosity and size on stellar mass, and the stellar mass -- halo mass relation implicit in the MAMBO lightcones, so we expect it to translate to real observations. This bodes well for studies of the most massive clusters with $\logMhalo > 15$.

The right panel of Fig.~\ref{fig:grid_rmax_mass} shows that $\Rmax$ increases as $\fICL$ increases. $\Rmax$ for the fiducial cluster ($\logMhalo=14.7$, $\fICL=0.15$, and $z=0.3$) decreases to only 550\,kpc if $\fICL = 5\%$, whereas we can measure the ICL out to over 900\,kpc if $\fICL = 25\%$.

\section{Discussion}\label{sec:discussion}
In this analysis, we have produced a set of mock \Euclid observations of clusters across a range of halo masses from $10^{13.9}$ to $10^{14.7}\,\Msun$, at redshifts $0.3 \leq z \leq 1.8$, and with ICL fractions ranging from $1\%$ to $25\%$. In Sect.\,\ref{sec:SNProfiles} we study the radial profile of the ICL S/N and motivate the adoption of a standard aperture (50--200\,kpc around the BCG) for comparing BCG+ICL fluxes between different clusters and different measurement methods.

We measured the total flux of the BCG+ICL \eceb{in the \Euclid \HE band}, and its uncertainty, using a straightforward aperture-photometry method and built up a grid of expected S/N for the \ICLap{} as a function of redshift, halo mass and ICL fraction. We explored the limits of this parameter space that we expect to be able to explore with \Euclid, considering both ICL S/N and the number of clusters of a given halo mass expected within the footprint of the entire EWS. We also tested several ICL modelling techniques to compare the resulting uncertainties with our theoretical limits. Finally, we investigated the outermost radial limits at which we expect to be able to detect ICL across the parameter space.

In summary, we expect that ICL will be detectable by \Euclid in clusters with $\Mhalo > 10^{14.2}\,\Msun$
up to $z \approx$ 1.3--1.6, depending on measurement uncertainties and assuming an optimal detection method. \Euclid\ will  be able to detect ICL fractions down to ${\sim}\, 1\%$ of the total light in clusters with $\logMhalo = 14.7$ up to $z=1.2$ ($\SN\sim3$).
These results are of importance to studies of ICL evolution: since simulations suggest that the ICL builds up over time, particularly in the later stages from $z = 0.4$ to $z = 0.1$ \citep{contini2021}, the expected limiting ICL fraction will give a realistic estimate of our ability to probe the important early stages of ICL formation. Currently, different studies show a large scatter (up to ${\sim}\,30\%$) in their predictions for the assembly of the ICL over time \citep{contini2024}. In \citet{furnell2021}, the ICL fraction was found to increase from ${\sim}\,10\%$ at $z=0.5$, to ${\sim}\,30\%$ at $z=0.3$. In \citet{burke2015}, a similar trend with redshift was found, however they measured significantly lower ICL fractions from ${\sim}\,2\%$ at $z=0.4$ to ${\sim}\,10\%$ at $z=0.3$.
In contrast, \citet{zhang2024} found no evolution of the ICL across the redshift range $0.2<z<0.5$.
Our expected sample sizes show that we can anticipate excellent coverage within this range of redshifts. \Euclid will be able to resolve the discrepancies between studies of the ICL fraction, exploring the growth of the ICL across a larger sample (up to ${\sim}\,19\,000$ clusters at $0.3 \leq z \leq 0.5$, see Fig.\,\ref{fig:nclust_SN}) and to higher redshifts, in order to constrain the growth of the ICL with cosmic time.

Whilst our results show that it will be challenging to detect ICL in clusters with $z>1$ with \Euclid, it will be a large improvement over other cosmological surveys with LSB capabilities (e.g., KiDS, DES, LSST), since these do not observe the NIR, and the optical bands only cover the rest-frame UV. Since ICL is very faint in the UV, the observations by \Euclid will make significant progress in this range.

In Sect.\,\ref{sec:extent}, we examined the ability of \Euclid to answer a variety of questions that can only be addressed by measurements of ICL at large radii.  These questions include the detection of the splashback feature \citep{gonzalez2021}, the characterisation of ICL spatial offsets from the BCG, and evolution in the ICL fraction.  Here, we present a brief description of what parameter space can be used to address each question.  

In both simulations \citep{deason21} and extremely deep observations \citep{gonzalez2021}, a splashback feature has been identified within the ICL.  \citet{gonzalez2021} identified a splashback feature in MACS J1149.5+2223, a massive cluster with $\logMhalo = 15.53$ at $z=0.544$, at a radius of 1.2--1.7\,Mpc, which is far beyond the parameter space (both radial range and $\Mhalo$) that we investigate in this paper.  However, as the splashback radius is a characteristic of the cluster, it likely scales with $R_{200}$ rather than lying at a fixed physical distance.  In simulations, \citet{deason21} estimate that this feature is at 0.7--1\,$R_{200}$.  This means that the splashback feature will only be observable in ICL with \Euclid in clusters more massive than $\logMhalo > 14.7$ at $z < 0.3$, or through stacking the ICL from many clusters with a certain richness or lensing mass.  

Recent observations of 170 clusters \citep{kluge20} have found that when the ICL is measured out to $\sim$\,200\,kpc, we are able to detect a moderate offset ($\sim30$\,kpc) between the center of the distribution of the ICL and the core of the BCG.  Such an offset was detected in the recent \Euclid ERO Perseus observations \citep{EROPerseusICL}. Assuming that the sample of \citet{kluge20}, $z\lesssim 0.08$; $14.3<\logMhalo<15.1$, is representative of the population to be detected in the EWS, and based on our sampled parameter space, we find that using \Euclid we should be able to characterise such offsets for all clusters with $\logMhalo > 14.2$ at $z < 0.9$ and some more massive clusters up to $z=1.2$.  Thus, this measurement should be attainable for the vast majority of clusters.  

Lastly, the ICL fraction is also commonly used to characterise how important a contributor the ICL is to the cluster's overall baryon budget.  Recent observations (done using stacked measurements) from \citet{zhang2024} have found that the ICL fraction varies depending on the selected radial aperture over which it is measured, decreasing by a factor of two when comparing the fraction in 30--80\,kpc to 300--600\,kpc.  Thus to accurately measure the ICL fraction requires measurements to be made to several hundred kiloparsecs. We estimate that such measurements could be made for all \Euclid-detected clusters with $\logMhalo > 14.5$ at $z < 0.3$ or all clusters $\logMhalo > 14.7$ at $z < 0.6$.

In Sect.\,\ref{sec:haloestimates}, we estimated the number of clusters in which we can measure the ICL across the extent of the EWS. We find that ICL can be detected in 80\,000 clusters of $\logMhalo>14.2$, half of which lie between $z=0.3$ and $z=0.75$, but only ${\sim}\,2000$ at $z>1.3$. These result indicate the expected sample size of the forthcoming \Euclid ICL survey. Making the most of these large predicted samples, stacking clusters will be an effective way of performing detailed studies using samples with lower individual S/N, as well as extending ICL measurements to higher redshifts. By stacking 50 individual clusters each with $\SN > 3$, a combined $\SN > 20$ can theoretically be achieved. Since we expect to observe ${\sim}\,2000$ clusters \eceb{beyond} $z=1.3$, most of which only have a $\SN\sim3$, stacking clusters will be an effective way of studying the behaviour of the ICL fraction at significantly higher redshifts than individual observations would permit. We note that these numbers can increase tenfold if ICL can be reliably detected in clusters of $\logMhalo<14.2$, without being affected by Eddington bias.  On the other hand, the cluster detection algorithm of \Euclid may not be able to identify all the clusters in which we are able to measure ICL, which would reduce the number of \Euclid clusters in which we can study the ICL. 

Prior to \Euclid, the Dark Energy Survey \citep[DES;][]{DES05} has been the only large cosmological survey with the capability of characterising the ICL in individual or stacked clusters \citep{zhang19, santos21, golden-marx23, golden-marx2024, zhang2024}.  However, DES's photometry only allows the ICL to be measured out to as far as $z=0.8$.  Therefore, the large sample sizes and expanded parameter space of \Euclid will be transformative for cluster evolution studies, particularly for tracing the build-up of the ICL over time.

\section{Conclusions}
\label{sec:conclusions}

In this paper, we have produced mock \HE-band images of galaxy clusters and ICL, across a range of halo masses ($13.9 < \logMhalo < 15.0$), redshifts ($0.3 < z < 1.8$), and ICL fractions ($0.01 < \fICL < 0.25$). We measured the radial profile of the ICL S/N, varying the aforementioned cluster properties, and defined a standardised aperture between 50--200\,kpc in which we score the success of an ICL detection.
We explored how the total S/N within the 50--200\,kpc annulus varies across the parameter spaces of halo mass, redshift, and ICL fraction: ICL will be detectable in \Euclid images with current methods in clusters with $\logMhalo \sim 14.2$ at $z<1.3$. In clusters with $\logMhalo > 14.4$, the ICL will be detectable up to $z=1.5$.

We investigated how these results translate to radial limits of ICL detection, both in terms of the limiting physical radius as well as the surface brightness limit. At $z=0.3$, we obtain a limiting radius of around 350\,kpc for a $\logMhalo \sim 14$ cluster, increasing to $\sim 800$\,kpc for a $\logMhalo \sim 14.7$ cluster. We found that the limiting surface brightness for an ICL fraction of 15\% and a halo mass of $\logMhalo=14.8$ ranges from 29\,mag\,arcsec$^{-2}$ at $z=0.3$ to ${\sim}\,28$\,mag\,arcsec$^{-2}$ at $z=1.2$.
Given the above predictions, we estimated the sample sizes of clusters with detectable ICL that will be attainable across the entirety of the EWS. Assuming an ICL fraction of $15\%$ and a S/N of 3, there exists ${\sim}\,80\,000$ clusters from $z=0.3$ to $z=1.6$ in which we could measure ICL, of which a couple of thousand will be at $z>1.3$. This number will likely reduce due to \Euclid's cluster selection function as well as regions contaminated by Galactic cirrus and bright stars, through which the ICL will not be visible.

In summary, we find that the EWS will provide opportunities to study the build up of the ICL with cosmic time and its formation mechanisms with an unprecedented sample of massive clusters.

\begin{acknowledgements}

NAH and JBGM gratefully acknowledge support from the Leverhulme Trust through a Research Leadership Award. CB, NAH, and SB acknowledge support from the UK Science and Technology Facilities Council (STFC) under grant ST/X000982/1. FD acknowledges support from CNES. This research was supported by the International Space Science Institute (ISSI) in Bern, through ISSI International Team project \#23-577.

\AckEC
\end{acknowledgements}

%
%

\bibliography{ref} 

%
%

\begin{appendix}
\begin{onecolumn}
\section{Impact of possible BCG contamination to ICL in 50--100\,kpc region}

\label{sec:contamination}

As discussed in Sects.~\ref{sc:Intro} and \ref{sec:SNProfiles}, separation of the BCG and ICL is challenging to do with only photometric information, and therefore we have treated the two as a continuous two-component profile in our S/N measurements. But this means that significant amounts of BCG light may fall in the 50--200\,kpc aperture we defined to measure the ICL S/N. In this section, we quantify the effect of the inner radius of the annulus on the ICL S/N measurements, since the inner radius dictates the contribution of the BCG to our ICL `signal'. 

With the aim of measuring the ICL in the lower mass clusters, this radius was set at 50\,kpc, as discussed in Sect.~\ref{sec:SNProfiles}, however this does introduce some contribution of the BCG to the total signal in the larger mass clusters. Figure\,\ref{fig:icl_bcg_ratio} shows that almost half of the light in this aperture may come from the BCG for the most massive clusters. We therefore explore using a narrower aperture of 100--200\,kpc around the BCG. In Fig.\,\ref{fig:icl_bcg_ratio}, we also plot the ICL/BCG ratio in the 100--200\,kpc aperture; in the lowest mass cluster, there is relatively less flux from the ICL component than the BCG (which has $n=4.6$) in the 100--200\,kpc annulus than the 50--200\,kpc, but for more massive clusters the ICL contribution is at least twice the BCG light contribution in this 100--200\,kpc annulus. This tells us that the S/N measured in this narrower aperture would be a purer measure of the S/N in the ICL for clusters with masses $\logMhalo>{14.2}$. 

In Fig.\,\ref{fig:mhalo_z_100_200}, we show the S/N measured within the 100--200\,kpc annulus as a function of redshift, halo mass and $\fICL$, which can be directly compared to the same results from the 50--100\,kpc annulus shown in Fig.\,\ref{fig:mhalo_z}. As expected, the overall S/N is lower, since there is less signal contributed by both the BCG and ICL in the narrower annulus. Clusters that were detected with a $\SN=10$ (5) in the 50--200\,kpc annulus would only be detected with a $\SN=5$ (3) in the 100--200\,kpc annulus. Thus, while a 100--200\,kpc annulus may provide a purer measure of the ICL, using this annulus for the ICL detection would reduce the number of detectable clusters to the quantities shown by the $\SN\geq5$ line in Fig.\,\ref{fig:nclust_SN}, thereby reducing the redshfit range over which ICL properties can be explored. 

\begin{figure*}[h!]
\centering
    \includegraphics[angle=0,width=0.92\hsize]{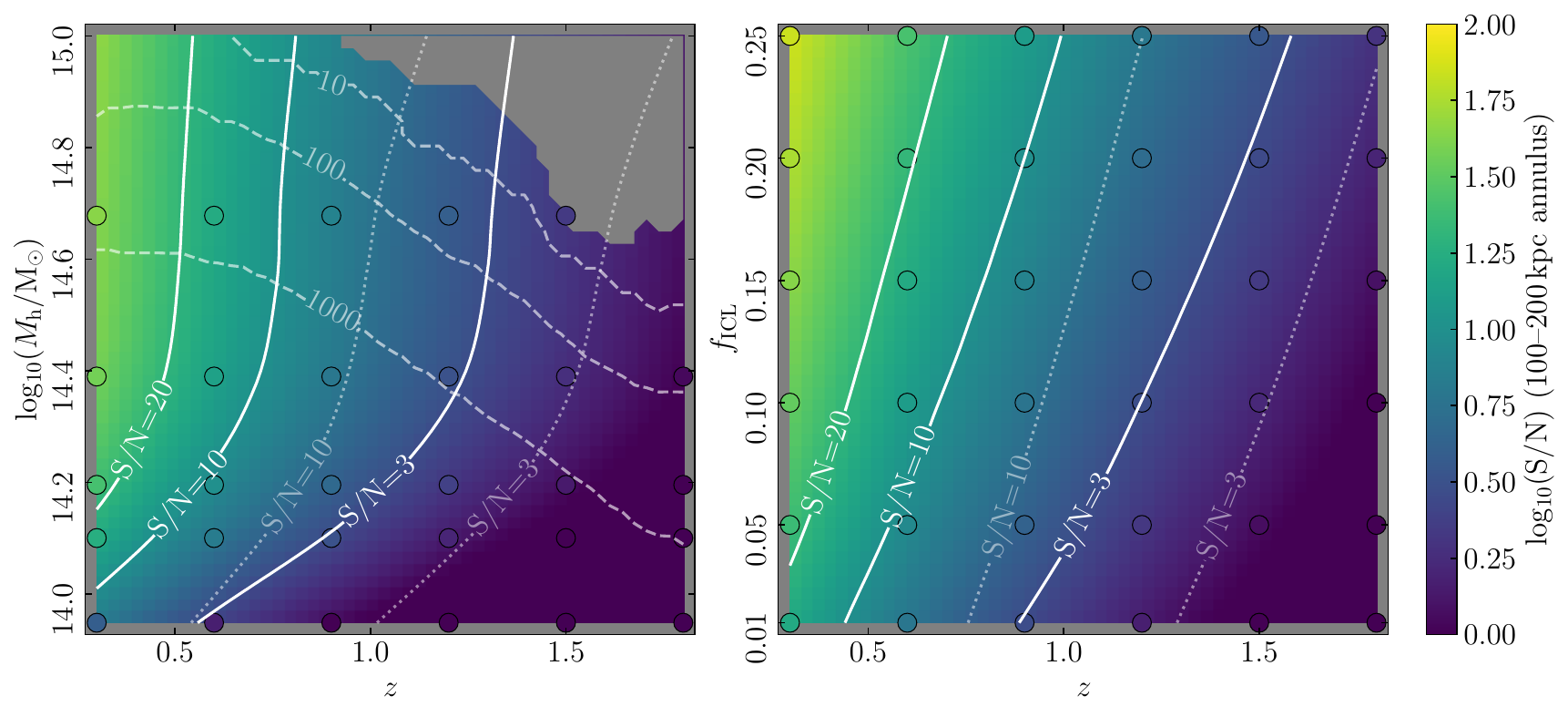}
\caption{\HE-band S/N of the ICL + BCG interpolated across halo mass, redshift space for a fixed ICL fraction of $15\%$ (left) and ICL fraction, redshift space at a fixed halo mass $\Mhalo=10^{14.7}\,\Msun$ (right). The figures are produced as described in Fig.\,\ref{fig:mhalo_z}, but for the S/N measured within a 100--200\,kpc annulus. All points and lines are coloured and drawn as described in Fig.\,\ref{fig:mhalo_z}, with the addition of two dotted lines showing the S/N=10 (left-most) and 3 (right-most) lines from the 50--200\,kpc annulus, for ease of comparison.}
\label{fig:mhalo_z_100_200}
\end{figure*}

\section{Modelling smooth variations in S/N and $\Rmax$ versus $\Mhalo$, $\fICL$, and $z$}
\label{sec:fitting_sn}

The cluster properties which dictate the flux and scale radius of our model ICL are the total cluster flux ($f_{\HE}$) and the cluster stellar mass ($\Mstar$) respectively. These values are dependent upon the individual clusters and have some inherent variation. We therefore modelled the total cluster flux and stellar mass as a function of halo mass across the sample of five clusters from MAMBO, in order to smooth out cluster-to-cluster variations.

For each cluster, we fit a linear function to the sum of the total \HE flux of all member galaxies as a function of total halo mass, $f_{\HE} = a\Mhalo$ where the best-fit values of $a$ are shown for each redshift in Table\,\ref{tab:slopes}. Then, at the mass of each cluster, we simulated ICL images using the interpolated value of the flux from this function rather than the measured value for the specific cluster, effectively smoothing out any differences between the clusters.
The same process was carried out for the stellar mass of each cluster, for which we found the best-fit equation: $\Mstar = 8.5\times10^{-3}\,\Mhalo$ for all redshifts.
We then used the interpolated values of cluster flux and stellar mass to set the flux of the ICL and BCG as well as the scale radius of the ICL.

\begin{table}[]
    \centering
    \caption{Best-fit slopes for the cluster flux ($a$) and galaxy noise ($b$) as a function of cluster halo mass ($\Mhalo$), for each value of $z$.}
    \label{tab:slopes}
    \begin{tabular}{c | c | c}
    $z$ & $a\,[10^{-12}\,\Msun^{-1}\,\si{\micro.Jy}]$ & $b\,[10^{-8}\,\Msun^{-1}\,\si{\micro.Jy}]$  \\\hline
    
    0.3 & 57 & 31\\
    0.6 & 11 & 15\\
    0.9 & 4.2 & 8.8\\
    1.2 & 1.9 & 5.8\\
    1.5 & 1.0 & 4.0\\
    1.8 & 0.60 & 2.7\\
    \end{tabular}
    
\end{table}

Differences in the luminosity and geometry of satellite galaxies introduce cluster-to-cluster variations in the noise. To account for this, we calculated the total noise contributed by cluster members within the 50--200\,kpc annulus ($\sigma_\mathrm{galaxies}$). We found that the total satellite noise within this region is well modelled by a function of the square root of the halo mass. This is expected given the total luminosity of the satellites scales approximately linearly with halo mass, and the Poisson noise scales with the square root of the flux. We modelled the galaxy noise as a function of the halo mass using the equation: $\sigma_\mathrm{galaxies}=b \sqrt{\Mhalo/\Msun}$, with the best-fit values of the slope $b$ shown in Table\,\ref{tab:slopes}.
We used this relation to regenerate the noise images, omitting the individual satellite galaxies, and measured the total noise of the remaining components (the BCG, ICL and mock observation noise) within the 50--200\,kpc annulus. When calculating the S/N, the interpolated value of the total satellite galaxy noise was added in quadrature to the total noise of the other components.

This method allowed us to extrapolate the flux, stellar mass, and satellite galaxy noise, in order to ascertain the values of those parameters at higher halo masses than our original sample of clusters permitted.
We extended the upper end of the mass range by simulating a cluster with halo mass $10^{15}\,\Msun$ approximately corresponding to the upper limit in halo mass of the sample used to derive Eq.\,\ref{eq:scaling_re}. Since the equation was modelled on a sample of galaxies ranging from $10^{14}$--$10^{15.1}\,\Msun$ \citep{kluge21}, extending significantly beyond these limits would likely introduce inaccuracies in our ICL model.

\section{Descriptions of ICL fitting methods}
\label{isopy_description}
\subsection*{Method 1: isophotal fitting of the BCG+ICL}

BCG+ICL models were created using the methodology described in \cite{Kluge2023} and \cite{EROPerseusICL}. In brief, we first measured the isophotal shapes and fluxes, and subsequently converted these parameters into 2-D image models. The isophotal shapes were measured by fitting ellipses to the lines of constant surface flux (isophotes) using the python package {\tt photutils} \citep{Bradley2023}. The median surface flux was then measured in elliptical annuli around these ellipses. This was done on masked images. The masks were generated by applying surface brightness thresholds on various spatial scales after modeling and removing larger spatial scales using 2-D splines. The procedure is detailed in \cite{kluge20} and the masking parameters were optimized for the simulated \Euclid images. Central sources near the BCG nucleus were masked by removing a first-estimate model for the BCG+ICL and remasking the residual image. The residual background constant was determined as the median surface flux of the five outermost data points after iteratively clipping data points that deviate by more than two median absolute deviations from the median. The outermost possible semimajor axis radius of 1.5\,Mpc was defined by the image cutout size of 3\,Mpc $\times$ 3\,Mpc. Finally, a 2-D image model of the BCG+ICL was generated by evaluating the interpolated surface flux profiles and isophotal shape profiles on a pixel grid.

\subsection*{Method 2: 1-D profile with satellite galaxy masking}

In this method, we first masked all the sources in the mock image field at each redshift with the nine cluster instances by running \textsc{SExtractor} \citep{bertin1996}. The \textsc{SExtractor} segmentation maps were radially extended by 10\,kpc before creating the masks to ensure that most parts of the diffuse light in the outskirts of the galaxies are excluded in our measurement. Then, we measured the overall background level for the field by taking the average value of the unmasked pixels with a $3\sigma$ clipping applied to the pixel values. To measure the BCG+ICL, we made cutouts of size 500\,kpc $\times$ 500\,kpc centred on each cluster. In each cutout, all the sources except for the BCG and the extended diffuse light are masked, and the azimuthally-averaged BCG+ICL surface brightness (SB) profile is measured in logarithmic circular apertures centred on the BCG. More details on the masking and SB profile measurement can be found in \citet{ahad2023}.

\subsection*{Method 3: wavelet fitting (DAWIS)}

DAWIS \citep{Ellien2021} is a prior-free multiscale de-noising algorithm that uses wavelet transforms to iteratively decompose an astronomical image into sets of 2D light components, denoted `atoms'. The sum of all atoms of an image results in a complete recovered astronomical field of all sources that have been detected, from bright foreground Milky Way stars to faint and extended sources such as the ICL. Complementary to this global modelling, one can separate the atoms based on their properties (such as spatial position and wavelet scale) to recompose the individual light profile of individual objects. Here DAWIS is used in such a way to produce BCG+ICL models of each mock cluster. First, smaller images of size $1500\times1500~\mathrm{pix}^2$ centred on each cluster BCG were extracted from all tiles to speed up computations (9 per tile). DAWIS is run on each cluster image individually with the same standard input parameters (see \citealt{Ellien2021} for a description). Upon finishing iterating, the atoms with light distribution centred on the BCG are summed for each image to produce a first estimate of the BCG+ICL light profile. A cut on the wavelet scale at which the rest of the atoms have been detected is applied (wavelet scale $\geq5$) to extract the leftover large-scale ICL signal, which is added to the BCG+ICL profile. This final 2D BCG+ICL light distribution is used to compute the fluxes in the 50--200\,kpc apertures as indicated in the main text.

\end{onecolumn}

\end{appendix}

\end{document}